\documentclass[12pt,preprint]{aastex}

\usepackage{graphicx}
\usepackage{txfonts}
\usepackage{longtable}
\usepackage{amssymb}
\usepackage{natbib}
\bibpunct{(}{)}{;}{a}{}{,}
\usepackage{subfigure}

\newcommand{\kms}{km~s$^{-1}$}
\newcommand{\Jyb}{Jy~beam$^{-1}$ }

\newcommand{\M}{CH$_3$OH}
\newcommand{\W}{H$_{2}$O}
\newcommand{\T}{W3(OH)}
\shorttitle{12~GHz maser kinematics in W3(OH)}
\shortauthors{Moscadelli et al.}

\begin{document}


\title{Revising the kinematics of 12~GHz \M\ masers towards \T}


\author{L. Moscadelli}
\affil{INAF, Osservatorio Astrofisico di Arcetri, Largo E. Fermi 5, 50125 Firenze, Italy}
\email{mosca@arcetri.astro.it}

\author{Y. Xu}
\affil{Purple Mountain Observatory, Chinese Academy of Sciences, Nanjing 210008, China}

\and

\author{X. Chen}
\affil{Shanghai Astronomical Observatory, Chinese Academy of Sciences, 80 Nandan Road, Shanghai 200030, China}




\begin{abstract}
We derive accurate proper motions of the \M\ 12~GHz masers towards the \T\ UC~H{\sc ii} region,
employing seven epochs of VLBA observations spanning a time interval of about 10~yr. 
The achieved velocity accuracy is of the order of 0.1~\kms, adequate to precisely measure the relative velocities 
of most of the 12~GHz masers in \T, with amplitude varying in the range \ 0.3--3~\kms. 
Towards \T, the most intense 12~GHz masers concentrate in a small area
 towards the north (the northern clump)
of the UC~H{\sc ii} region.
We have compared the proper motions
of the \M\ 12~GHz masers with those (derived from literature data) 
of the OH~6035~MHz masers, emitting from the same region of the methanol masers.  
In the northern clump, the two maser emissions emerge from nearby (but likely distinct) 
cloudlets of masing gas with, in general, a rather smooth variation of line-of-sight and sky-projected velocities,
which suggests some connection of the environments and kinematics traced by both maser types.
The conical outflow model, previously proposed to account for the 12~GHz maser kinematics
in the northern clump, does not reproduce the new, accurate measurements of 12~GHz maser proper motions and has to
be rejected. We focus on the subset of 12~GHz masers of the northern clump belonging to the
``linear structure at P.A. = 130\degr--140\degr'',
whose regular variation of LSR velocities with position presents evidence for some ordered motion.
We show that the 3-dimensional velocities of this ``linear distribution'' of 12~GHz masers
can be well fitted considering a flat, rotating disk, seen almost edge-on.
\end{abstract}





\section{Introduction}

\label{intro}

Studying the evolution of  H{\sc ii} regions and their complex interaction with the
surrounding environment, is important in the context of star formation. H{\sc ii} regions
emit intense far-ultraviolet (FUV) radiation
which photoionizes the circumstellar gas and evaporates more volatile molecules frozen on dust grain mantles.
The physical and chemical properties of the gas in the natal molecular core (with typical size of about 0.1~pc)
are greatly altered, with a major impact on the subsequent star-formation conditions.
The radiation pressure exerted by the H{\sc ii} regions onto dust grains can eventually significantly contribute to the disruption
of the molecular core, too.

Hyper-compact (HC) and ultra-compact (UC)  H{\sc ii} regions are clear indicators of forming (or recently born) massive stars.
In regions of massive star-formation 
are often observed intense maser emissions of OH (at frequency of 1.6~and~6.0~GHz) , \W\ (22~GHz)
and \M\ (6.7~and~12~GHz).  Using multi-epoch Very Long Baseline Interferometry (VLBI) observations,
one can accurately determine absolute positions and velocities of maser spots (the single maser
emission centers), providing unique information on the kinematics of the molecular gas
around massive Young Stellar Objects (YSO).

Learning from the few sources studied in details so far, it appears
that \W\ 22~GHz and OH 1.6~GHz masers could be tracing respectively  the fast (50--100~\kms)
and slow (5~\kms) expansion of  compact H{\sc ii} regions in different evolutionary stages \citep{Tor03, Mos07, Fis07b}.
Accurate measurements of \emph{internal} motions of \M\ 6.7~and~12~GHz masers are still lacking in the literature.
In several sources (single-epoch) VLBI observations have shown linear or arc-like distributions
of maser spots, which have been interpreted in terms of edge-on rotating toroids or disks \citep{Nor98,Pes05,Pes09}.
However, in a few cases such elongated maser distributions lie in projection on the sky
parallel to typical outflow tracers (such as the H$_2$ 2.12$\mathrm{\mu}$m line), rather suggesting association with
outflowing gas \citep{Deb03}. Measuring \emph{relative} proper motions of \M\ maser spots appears the most direct way to discriminate
between rotation and outflow.

\T\ is one of the best studied UC~H{\sc ii} region. It is associated with a far-infrared source with a
total luminosity of 10$^5$~L$_{\sun}$ \citep{Cam89}, which corresponds to a main-sequence O7 star.
It harbours a plethora of maser transitions, among which the most intense and best studied by means of VLBI
are those of OH at 1.6 \citep{Blo92,Wri04,Fis06}, 6.0 \citep{Des98, Fis07a} and 13~GHz \citep{Bau98}, and
those of \M\ at 6.7 \citep{Men92} and 12~GHz \citep[][hereafter MMWR1, MMWR2 and XRZM, respectively]{Mos99, Mos02,Xu06}.
OH 1.6~and~6.0~GHz and \M\ 6.7~GHz masers are distributed across an area of about 2\arcsec in size, covering all
the western half of the continuum emission, where the H{\sc ii} region is confined by relatively denser molecular gas.
In contrast, most of \M\ 12~GHz maser emission comes from a small (diameter of about 200~mas) cluster (the ``northern clump'')
 to the north of the UC H{\sc ii} region \citep[][Fig.~3]{Mos99}, in correspondence with the northern ionized clump observed in high frequency
(15--23~GHz) continuum images of \T. The highly excited OH 13~GHz masers also emerge from a compact region
(extended about 100~mas) of the northern ionized clump. This area, including highly excited maser transitions,
the brightest continuum emission, and  showing also intense magnetic fields \citep{Wri04}, is the most active of the whole UC H{\sc ii} region and
could host the main source of excitation in \T. 

By comparing two Very Long Baseline Array (VLBA)\footnote{The VLBA is
operated by the National Radio Astronomy Observatory (NRAO).}
epochs separated by about 5~yr, MMWR2
first measured relative proper motions of \M\ 12~GHz masers in \T\
and found typical velocity amplitudes of a few \kms.
Positions and velocities of the 12~GHz masers in the northern clump
have been fitted with a narrow conical outflow model oriented
at close angle with the line of sight. However, most of the measured proper motions
have large ($\ga$50\%) uncertainties and the outflow model was effectively constrained by the (well-known)
maser line-of-sight velocities only. Recently, by using 5 VLBA epochs,  XRZM have measured the parallax of the
\M\ 12~GHz masers in W3(OH), deriving a very accurate source distance of $1.95\pm0.04$~kpc. 
To better constrain the kinematics of the 12~GHz masers, in this work we combine the MMWR2 and XRZM
12~GHz maser observations (forming a dataset of seven VLBA epochs spanning about 10~yr) and
derive accurate, relative proper motions for the persistent maser spots.

Section~\ref{sum} of this paper briefly describes the VLBA observations of 12~GHz methanol masers which we employ for proper motion
derivation. Section~\ref{var} presents a study of the 12~GHz maser variability over the 10~yr period spanned by the VLBA
observations. In Sect.~\ref{met_pm} we derive the 12~GHz maser proper motions and compare the new values with the
previous measurements by MMWR2. Basing on two published VLBI observations of OH 6.0~GHz masers in \T,
Sect.~\ref{oh_pm} derives proper motions of this maser transition and compares the velocity distributions
of OH 6.0~GHz and \M\ 12~GHz masers in the  northern clump.  Finally,  in Sect.~\ref{mod_kin},
the conical outflow model by MMWR2 is tested against the more accurate 12~GHz maser proper motions
derived in Sect.~\ref{met_pm}. Conclusions are drawn in Sect.~\ref{conclu}.


\section{Summary of \M\ 12~GHz maser VLBA observations}

\label{sum}

Table~\ref{obs_vlba} reports the main parameters of  our 12~GHz maser VLBA observations.
A full description of the VLBA observational setup, data calibration and analysis is given in MMWR1, MMWR2 and XRZM.
The observations span a time interval of about 10.5 yr. The XRZM five epochs are distributed over one year (from July 2003  to July 2004)
for the purpose of measuring the parallax of the 12~GHz masers.
Tables~2 of MMWR1 and Tables~1 of MMWR2 report spot parameters (position, intensity, FWHM size)
derived by fitting a two-dimensional Gaussian  profile to the spot intensity distribution.
In these tables, the ``Component'' label given in Col.~1 identifies the maser spots persistent across the two epochs.
In this work, we will name maser spots using the same ``Component'' labels of MMWR1 and MMWR2.
Position offsets given in Tables~2 of MMWR1 and Tables~1 of MMWR2, are relative to the maser spot \#~23,
which was used as phase-reference in the analysis of the first epoch (1994) VLBA data.
Maser spot parameters  were derived by MMWR1 and MMWR2 on images hanning-smoothed to a velocity resolution of 0.5-0.6~\kms.
With such a resolution, most of the spot emission is contained in a single velocity-resolution element.

\begin{figure}
   \centering
   \includegraphics[angle=0,width=9cm]{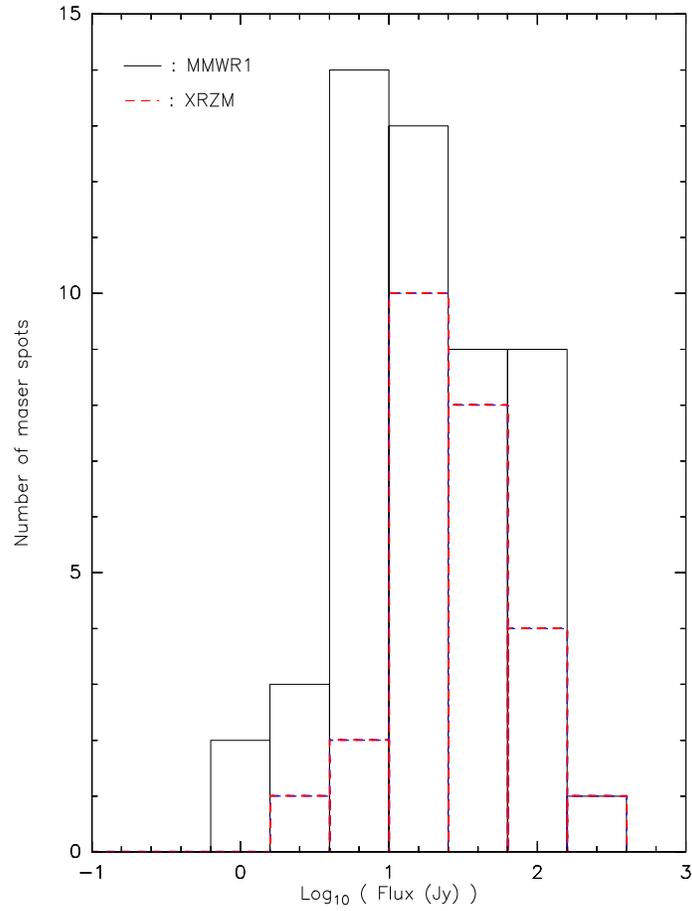}
      \caption{The histograms of the fluxes of maser spots detected by MMWR1 and XRZM are shown
                       in \emph{black solid} and \emph{dashed red line}, respectively. The $x$-axis is logarithmic and the histogram
                      bin is 0.4 dex.
              }
         \label{histo_flux}
   \end{figure}

XRZM determined spot parameters from images with a velocity resolution of 0.4~\kms,
comparable with that of the hanning-smoothed images of MMWR1 and MMWR2. Table~\ref{obs_vlba}
shows that also the angular resolution of the seven VLBA epochs was similar.
Thus, the spot parameters (from a two-dimensional Gaussian fit) derived by XRZM can be directly compared with
those of MMWR1 and MMWR2. It is to be noted that XRZM observations were less sensitive than the MMWR1~and~MMWR2 VLBA epochs.
Figure~\ref{histo_flux} shows the distribution in fluxes of maser spots detected by MMWR1 and XRZM.
Only 3 out of the 19 spots with flux less than 10~Jy observed by MMWR1, were recovered by XRZM.
However, the sensitivity of the XRZM VLBA observations was sufficient to recover most of the spots stronger than 10~Jy with good signal-to-noise
ratio, allowing an accurate determination of their proper motions.

\clearpage
\begin{table}
\caption{\M\ 12~GHz VLBA Observations}             
\label{obs_vlba}      
\centering                          
\begin{tabular}{c c c c c c}        
\tableline\tableline                 
\noalign{\smallskip}
Epoch & Date & \multicolumn{2}{c}{Uniform. Beam} & Vel. Res. & Ref. \\    
           &         &  Maj. Axis & Min. Axis &   &  \\
          & (yr m)  & (mas) & (mas) &  (\kms) &      \\
\noalign{\smallskip}
\tableline                        
\noalign{\smallskip}
   1 & 1994 February  &  1.1 & 0.7  & 0.3 & 1  \\      
   2 & 1998 December & 1.1 & 0.7 &  0.02 & 2   \\
   3 & 2003 July & 1.4 & 0.9 &  0.4 & 3      \\
   4 & 2003 October & 1.4 & 0.9 & 0.4 & 3     \\
   5 & 2004 January & 1.4 & 0.9 & 0.4 & 3     \\
   6 & 2004 April & 1.4  & 0.9 & 0.4 & 3     \\
   7 & 2004 July & 1.4 & 0.9 & 0.4 & 3     \\
\noalign{\smallskip}
\tableline   
\noalign{\smallskip}
\end{tabular}
\tablecomments{ References: (1) MMWR1; (2) MMWR2 ; (3) XRZM.}
\tablecomments{ \footnotesize  Columns~1~and~2 give
the VLBA epoch and the observing date, respectively; Cols.~3~and~4 report the FWHM Major and Minor axis
 of the uniformly-weighted beam used to reconstruct the maser images; Col.~5 indicates the velocity resolution of the
correlated visibilities; Col.~6 gives the references.}
\end{table}

%


\section{Time Variability}

\label{var}

\begin{figure*}
   \centering
   \includegraphics[width=14cm]{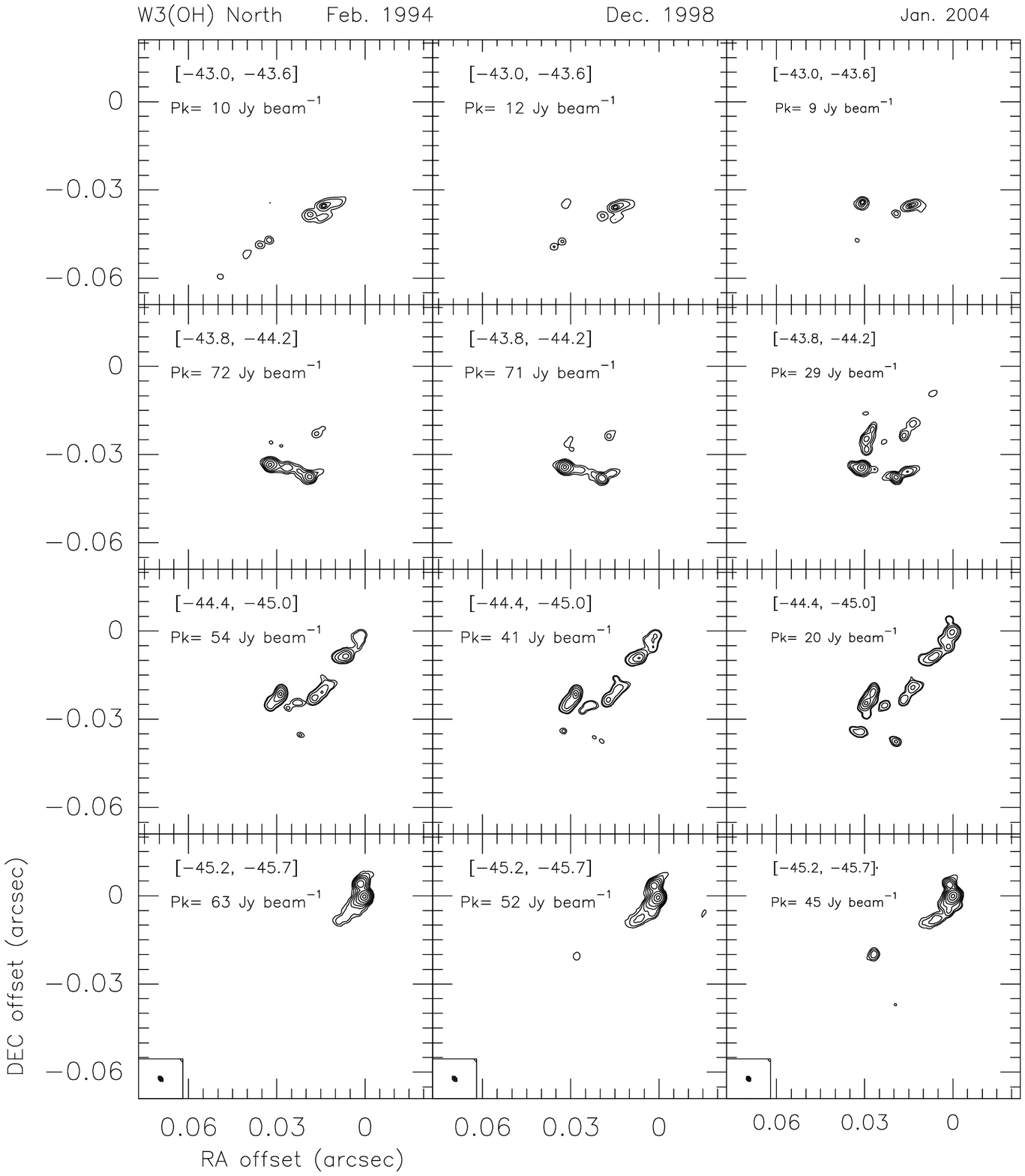}
      \caption{\small Left, middle and right panels present the velocity-integrated emission of \M\ 12~GHz
               masers in the northern clump, at the first (1994 February), second (1998 December),
               and fifth (2004 January) VLBA epoch, respectively. All panels show the same field of view, with
               positions referred to the 1994 phase-reference maser spot (\#~23).
               On the upper-right corner of each panel, the averaging velocity interval (in \kms)
               and the intensity peak of the image, are reported. Panels on a given row compare emission
               integrated over the same velocity interval, and plotted using the same peak percent levels.
               Plotted levels are 15, 30, 60, 80, 90\% of the peak for the first-row panels;
                 4, 8, 15, 30, 60, 90\%  for the second-row panels;
                 5, 7, 15, 30, 60, 90\% for the third-row panels;
                 1, 2, 4, 8, 15, 30, 60, 90\% for the fourth-row panels.
                 In each of the three epochs, the maser images have been restored using the same beam
              ($1.1 \times 0.7$~mas, P.A. = $8\degr$), shown in the inset on the bottom-left corner
               of the last-row panels.
              }
         \label{int-map_N}
   \end{figure*}

Figure~\ref{int-map_N} compares the velocity-integrated emission of  \M\ 12~GHz masers in the northern clump
among the 1994 February,  1998 December and  2004 January  VLBA epochs. The plotted field of view corresponds to an
area of about 100~mas in size, where the strongest 12~GHz masers in \T\ are clustering. The VLBA at 12~GHz
achieves an angular resolution sufficient to partially resolve and determine the maser structure, which appears as a collection of
bright spots connected by a bridge of more extended and weaker emission.

By inspecting panels showing emission
at corresponding velocity intervals, it is possible to identify bright maser spots persisting over the epochs
at similar relative positions. While at the extreme velocities
the maser intensity keeps approximately constant or decreases slightly,
at the central velocities it diminishes by a larger factor of 2.5--3.
It is remarkable the regular variation over the epochs of the relative intensity
of the weaker emission features, which stably become either brighter or dimmer.

In the second VLBA epoch (1998 December) 50 out of the 51 spots detected in the first epoch (1994 February) were recovered.
MMWR2 note that, even if relative spot intensities can vary by factors up to 3--4, the maser emission structure
remains remarkably similar across about 5~yr. In the last five VLBA epochs (2003-2004), XRZM recover 26 out of the
51 spots detected in the first VLBA epoch. Most of the strongest maser spots (with intensity higher than 5~\Jyb) persist over a time span of about 10.5~yr.
The 25 spots missed in the last VLBA epochs had intensities of only a few~Jy  in the first~and~second-epoch  maps,  and their non detection can be
ascribed to the lower sensitivity of the XRZM VLBA runs and also to maser variability.
Three weak spots, not previously observed by MMWR1 and MMWR2, are new detections by XRZM. Since the observations by XRZM  cover a time span of  only 1~yr,
too small to derive accurate relative proper motions, in our analysis we will not consider these new detections further.

%


\section{\M\ 12~GHz maser proper motions}

\label{met_pm}

   \begin{figure*}
    \centering
   \subfigure[{\small
The left and right panel presents the linear fit of the change with time
 of the spot spatial coordinate towards the east and the north, respectively.
    \emph{Errorbars} indicate spot positions (and the associated errors) relative to the reference spot \#~23.
    Each \emph{black line} gives the best linear fit for the motion of the spot identified
    by the ``Component'' label  (see Table~\ref{prmot}) reported on the upper left side of the fit.
    In each panel, the vertical bar showes the amplitude scale of spot position offsets.}
       ] 
{
    \label{prmot_fit_a}
    \includegraphics[width=7.4cm]{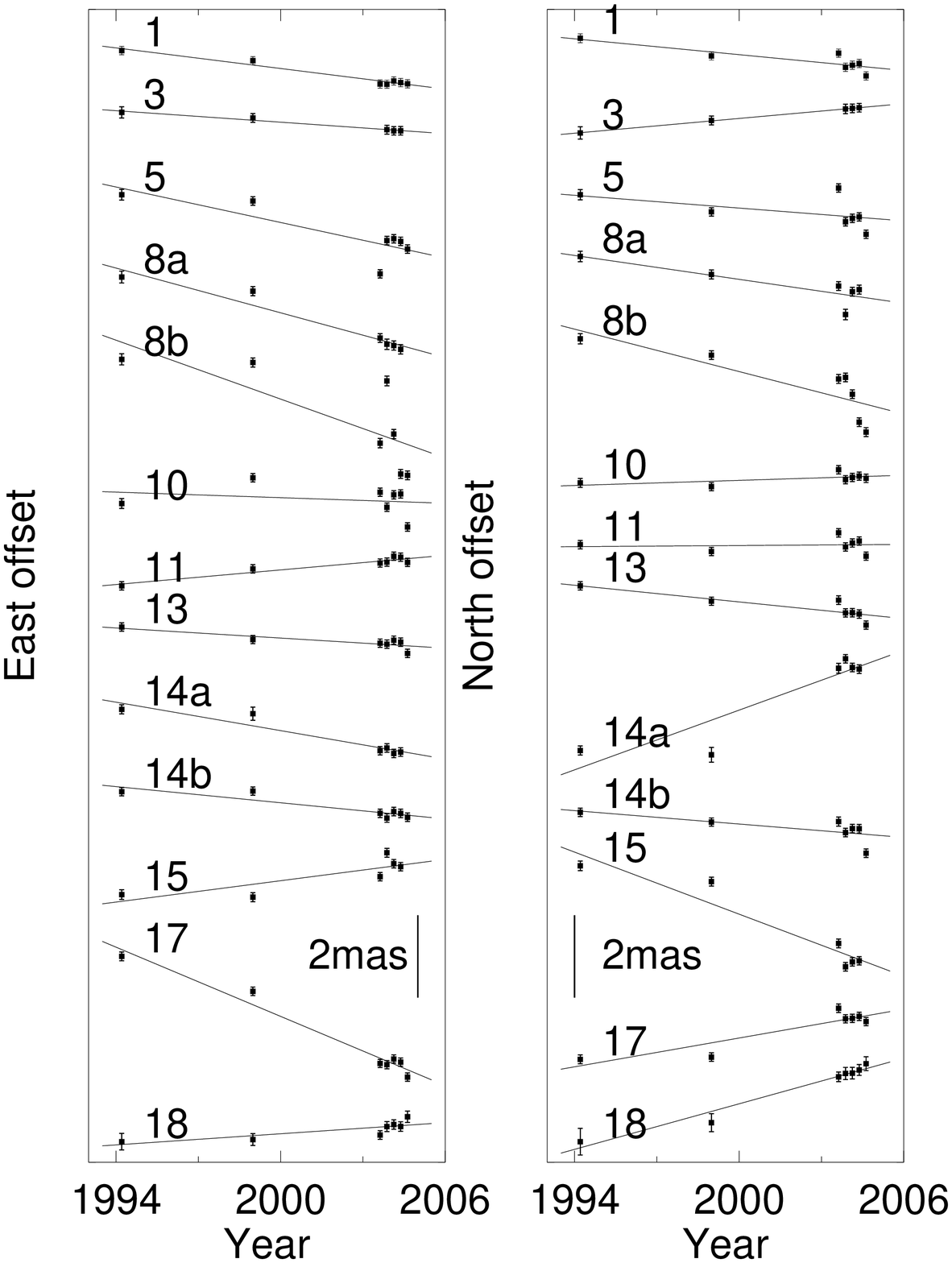}
}
\hspace{1cm}
\subfigure[Same as for (a)] 
{
    \label{prmot_fit_b}
    \includegraphics[width=7.4cm]{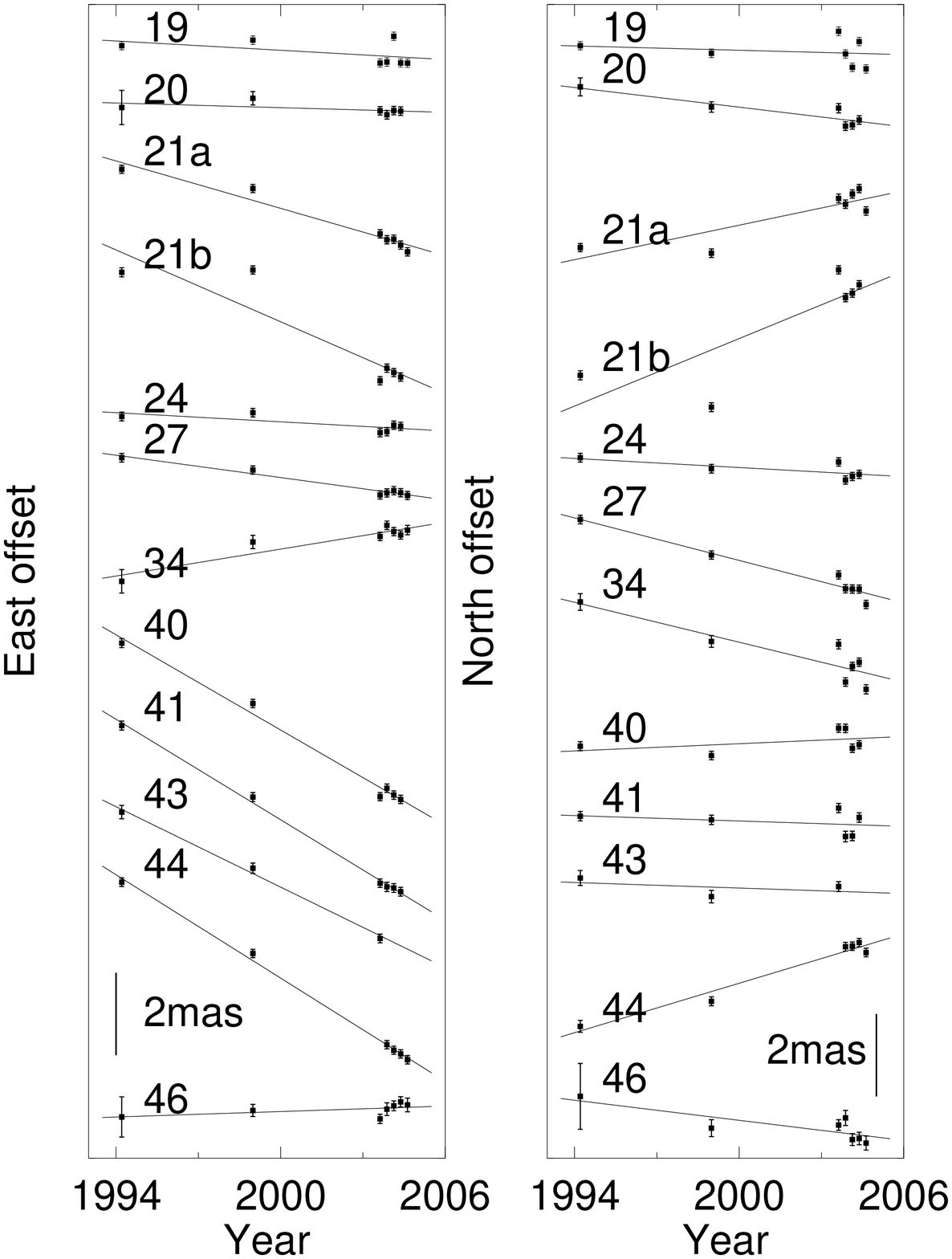}
}

\caption{Proper motion fit for spot persistent over three or more VLBA epochs.}
\label{prmot_fit}
   \end{figure*}

Table~\ref{prmot} lists proper motions of \M\ 12~GHz masers in \T,
derived for spots which persisted over a time span equal or larger than about 5~yr
(the separation between the first and second VLBA epochs).
Column~1 identifies the maser
spot reporting the same ``Component'' label used by MMWR1 and MMWR2.
Three spots with complex emission are fitted with two spatially-blended components,
denoted using the same integer number and lowercase letters ``a'' and ``b''.
Proper motions are relative to the maser spot \#~23, to which maser positions were referred
at each VLBA epoch. Proper motions derived for spots observed over the first two VLBA epochs only,
are, on average, significantly more uncertain than those calculated with three
 or more VLBA epochs\footnote{Since we do not consider the few new detections by XRZM, spots observed over three or more VLBA epochs have been
detected at both first (1994) and second (1998) VLBA epoch plus at least one of the XRZM (2003--2004) epochs, and therefore are persisting over a time baseline of about 10~yr.}
over an about double time baseline. Estimated errors of two-epoch proper motions are generally larger than 0.3~\kms,
whereas the best determined proper motions have uncertainties as small as 0.07~\kms.

The proper motions listed in Table~\ref{prmot} are derived by means of a linear fit of spot relative position vs time.
Figure~\ref{prmot_fit} shows the proper motion fit for spots persistent over three or more VLBA epochs.
Note that for most spots, the assumption of a uniform motion across the VLBA epochs fits well the observations.
The spatially-blended components ``8b'', ``14a'' and ``21b'', present the largest departure from linear motion.
In MMWR1 and MMWR2, spot position errors were calculated with the standard formula for the Gaussian fit:
$ 0.5 \, \mathrm{FWHM} / \mathrm{SNR} $, where FWHM is the fitted spot size and SNR is the ratio of the spot intensity with the
image rms noise. Since 12~GHz maser emission in \T\ is partially resolved with VLBA, change in the structure of the
phase-reference spot and maser spatial blending usually dominate the error-budget over signal-to-noise limitations.
It is difficult to estimate this error contribution, but our experience is that it can vary from case to case
in the range 10-100~$\mathrm{\mu}$as for isolated spots, increasing up to several hundreds of microarcsec for
more spatially-blended components.
To calculate more reliable spot position errors, we added a constant error floor of 50~$\mathrm{\mu}$as
in quadrature with the formal errors of the Gaussian fit. Position errors listed in Table~\ref{prmot} and position errorbars shown in Fig.~\ref{prmot_fit}
have been derived following this rule. For spatially-blended components, more heavily affected by systematic errors of Gaussian fit,
the calculated position errors are likely underestimated.
While for isolated spots we rely on the estimated values of position errors to derive proper motion errors with the linear fit,
for spatially-blended spots we adopt a different strategy.
For these spots, proper motion errors have been estimated
from the rms of residuals of the linear fit of spot position vs time,
and the original (a priori) position errors have been rescaled by the factor 
required to make their rms equal to the rms of fit residuals.

Table~2 of MMWR2 reports \M\ 12~GHz maser proper motions derived by maximizing the cross-correlation of the intensity distribution
of corresponding spots between the 1994 February and 1998 December VLBA epochs. In principle, the proper motion of extended objects,
assuming the internal structure does not change significantly with time, can be better determined by cross-correlating the emission distribution than
simply considering the motion of the emission peak. Comparing the velocities of Table~\ref{prmot} with those of Table~2 of MMWR2,
we found out that the velocity component towards the north was derived with the wrong sign by MMWR2.
In the procedure of calculating the cross-correlation function of spot intensity distribution, the conversion to velocity of the correlation steps towards the north,
expressed originally in terms of map pixels, was correct in absolute value but had the wrong sign.

  \begin{figure*}
   \centering
   \subfigure
{
    \label{prmot_comp_x}
    \includegraphics[width=7cm]{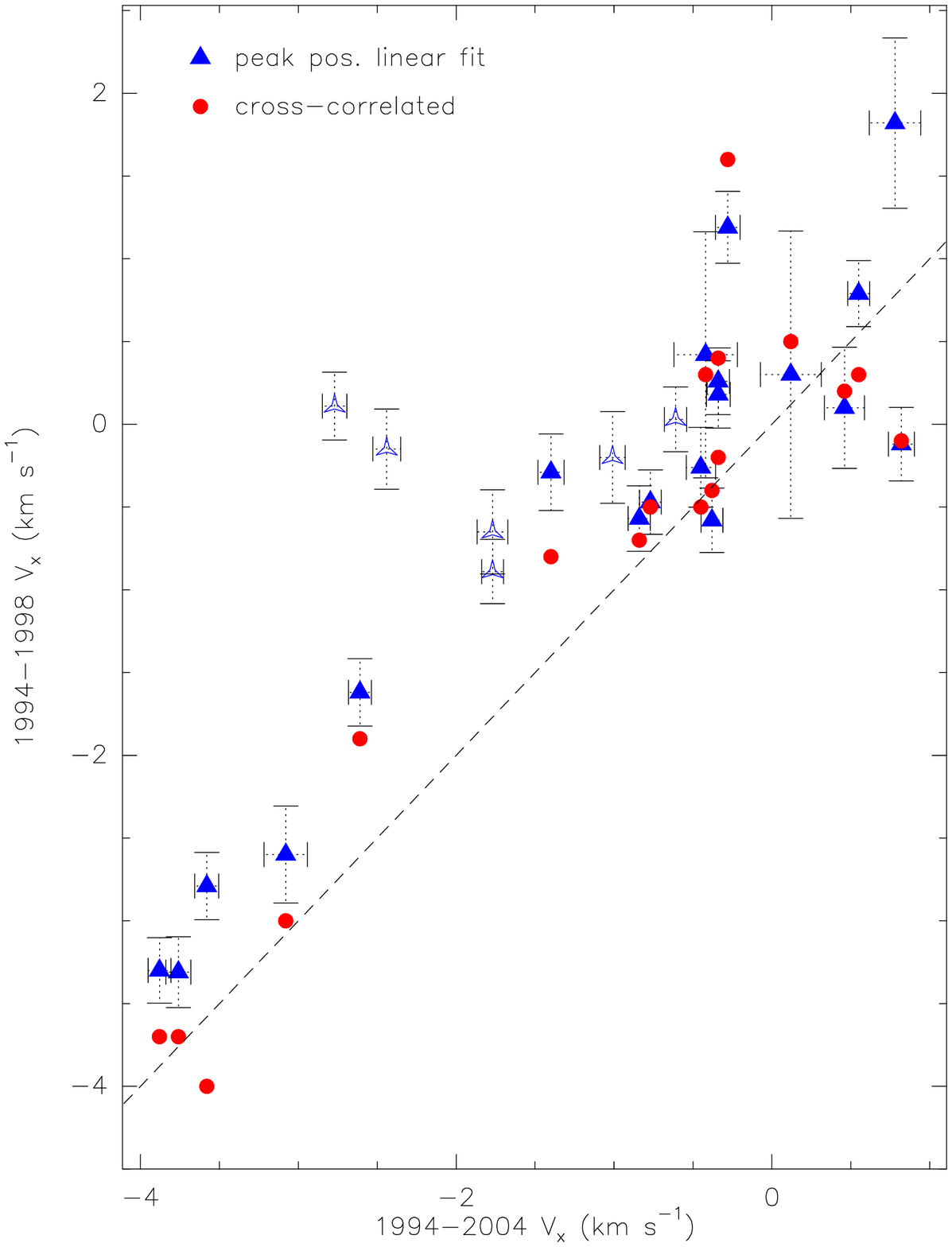}
}
\hspace{1cm}
\subfigure
{
    \label{prmot_comp_y}
    \includegraphics[width=7cm]{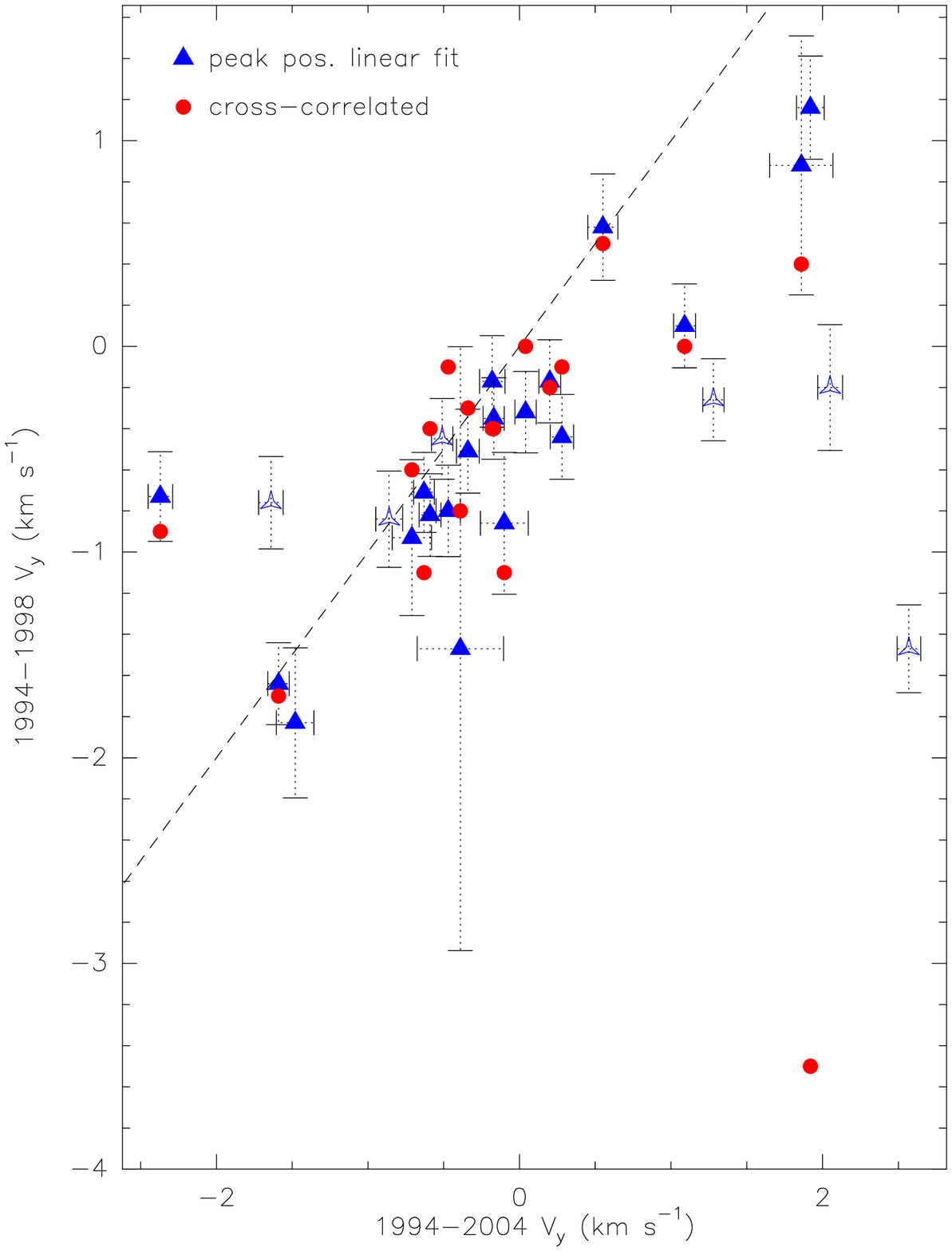}
}
\caption{Comparison of proper motions measured using the two first VLBA epochs only,
                    with values derived using three or more VLBA epochs.
                 Left and right panel report velocity component towards the east and north, respectively.
                 The horizontal axis reports measurements obtained using all the epochs of detection of a given spot,
                 whereas values on the vertical axis result from data of the first two epochs only.
                Proper motions over three or more epochs are derived by performing
                a linear fit of the change with time of the spot peak position.
               Two different measurements of two-epoch proper motions are shown:
               1) by maximizing the cross-correlation of the extended spot emissions;
               2) by simply considering the displacement of the spot emission peak between the epochs (the linear-fit method).
               \emph{Blue triangles} and \emph{red dots} show values derived with the linear-fit and cross-correlation, respectively.
                \emph{Empty triangles} denote spatially-blended double components,
                whilst \emph{filled triangles} are used for more isolated spots. \emph{Vertical} and \emph{horizontal errorbars} give velocity errors
               for proper motions calculated with two or more-than-two VLBA epochs, respectively.
                In both panels the \emph{dashed line} represents the \, Y = X \, line.}
\label{prmot_comp}
   \end{figure*}

Selecting the spots observed at three or more epochs, Figure~\ref{prmot_comp} compares proper motions calculated using the first two
VLBA epochs only, with the values derived using all epochs of detection of a given spot.
Two different measurements of two-epoch proper motions are shown: 1) values from Table~2 of MMWR2 (reversing the sign of the velocity component towards the north),
obtained via cross-correlation of extended spot emission; 2) by simply considering the displacement of the spot emission peak between the epochs (the linear-fit method).
Apart from a few spatially-blended spot components, whose values are more scattered, proper motions measured with two VLBA epochs over a time baseline
of 5~yr are consistent with measurements obtained with three or more VLBA epochs over a time baseline of 10~yr. Excluding the double components, the error-weighted
rms difference between linear-fit measurements using two and more-than-two epochs, is 0.7~and~0.6~\kms\ for the velocity component towards the
east and north, respectively. Using the two-epoch proper motions derived via cross-correlation of spot emission, the error-weighted rms difference
is 0.4~and~0.7~\kms\ for the velocity component towards the east and north, respectively. The fact that the spatially-blended spot components
deviate from a linear motion more than isolated spots, suggests indeed that their Gaussian-fit position is affected by larger uncertainties.

Figure~\ref{prmot_comp} suggests that, for spots of sufficiently simple structure, two VLBA epochs separated by about 5~yr allow to measure
\M\ 12~GHz maser proper motions towards \T\ with an accuracy of 0.5--1~\kms. Table~\ref{prmot} reports values of proper motions (in italic characters)
also for spots observed during the first two VLBA epochs only. Velocity errors of most of these spots, as determined by the linear fit,
 vary in the range 0.2--0.5~\kms, and are possibly underestimated by a factor of 2--3.  Taking the most accurate proper motions
of Table~\ref{prmot}, 12~GHz maser internal motions in \T\ should be mainly in the range 0.3--2~\kms. Thus, proper motions
determined with the first two VLBA epochs are likely affected by large relative errors, from 30\% up to 200\% for the slowest spots. On the other hand,
the accuracy achieved
with several VLBA epochs over a time baseline of about 10~yr is of the order of 0.1~\kms\ and appears adequate to measure 
with an accuracy better than 30\% most of  the 12~GHz maser proper motions in \T.

  \begin{figure*}
   \centering
   \includegraphics[width=\textwidth,angle=-90]{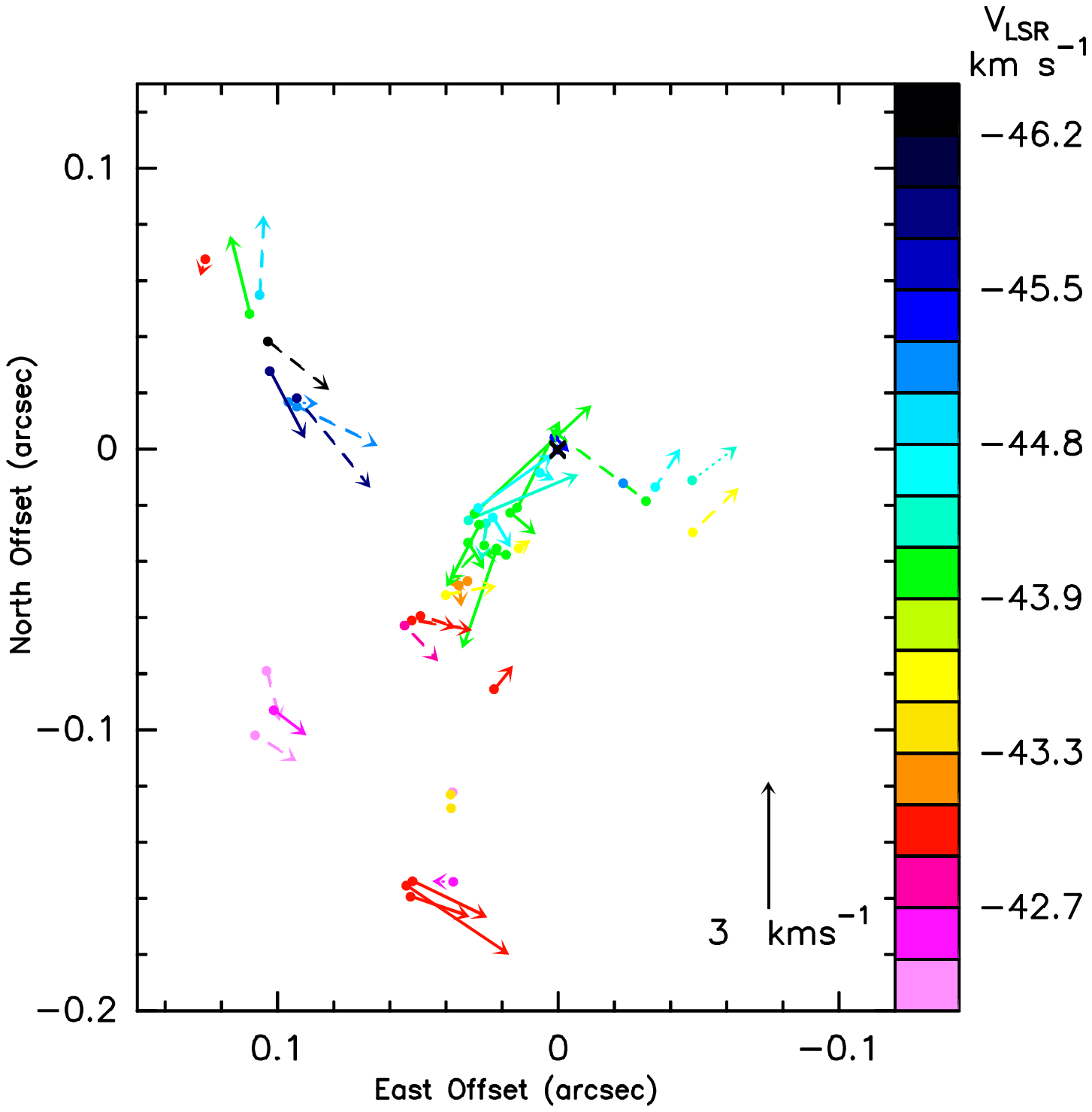}
      \caption{Spatial distribution and relative proper motions of \M\ 12~GHz masers in the northern clump of \T.
      \emph{Dots} give spot positions relative to the phase-reference spot  (``Component'' \#~23 in Table~\ref{prmot}) marked with a cross.
      \emph{Colors} are used to indicate the maser LSR velocities, according to the colored scale on the right-hand side
       of the plot, with \emph{green} denoting the  mean LSR velocity, at $-44.1$~\kms, of persistent \M\ 12~GHz masers.
      \emph{Arrows} indicate proper motions relative to the phase-reference spot, with \emph{dashed} and \emph{solid arrows} denoting proper
      motions measured using two or more-than-two VLBA epochs, respectively.  \emph{Dotted arrows} are used for the most uncertain measurements.
      The proper motion amplitude scale is given at the bottom right of the plot.
              }
         \label{prmot_rel_N}
   \end{figure*}

Figure~\ref{prmot_rel_N} shows the spatial distribution and the relative proper motions 
of the 12~GHz masers in \T\ belonging to the northern clump.
The reference spot used to refer maser positions and velocities is marked with the black cross. Note that clusters of nearby spots tend to show
similar line-of-sight and sky-projected velocities, which reinforces our confidence on the reliability of the derived proper motions.
One has to keep in mind that those shown in Fig.~\ref{prmot_rel_N} are velocities \emph{relative} to one spot chosen as reference, whose
motion with respect to the UC~H{\sc ii} region and/or the central ionizing star is unknown.
Looking at the global distribution of maser velocities in the northern clump, there is no evidence for a regular variation of velocity with position.

%


\section{OH 6.0~GHz maser proper motions}

\label{oh_pm}

\citet{Des98} and \citet{Fis07a} have used the European VLBI Network (EVN)
 to observe the excited OH maser transitions at frequencies of 6031~and~6035~MHz towards \T.
Most of maser spots observed by \citet{Des98} in 1994 May were recovered by \citet{Fis07a} in 2006 June.
Comparing the two EVN epochs separated by a quite long time baseline of about 12~yr, allows one to derive accurate values
of OH maser proper motions.
\citet[][Fig.~4]{Fis07a} show the distribution of the proper motions of OH 6031~and~6035~MHz masers across the whole extent of the W3(OH)
UC~H{\sc ii} region. In the following we present a more detailed comparison of  OH and \M\ maser velocities
in the northern clump of \T.

\citet[][Tables~1~and~2]{Des98} and \citet[][Table~1]{Fis07a} report
the relative positions of OH 6031~and~6035~MHz masers (for both LCP and RCP polarization)
measured with EVN in 1994 May and 2006 June, respectively.
To establish spot correspondence between the two EVN epochs, we follow the indications
by \citet{Fis07a} who point out which are the two spots of the 2006 observations corresponding to
the 6031~and~6035~MHz phase-reference spots of the 1994 observations. Examining relative
positions, it is then possible to unambiguously establish the correspondence between the 2006- and 1994-epoch
detections.
We correct the LSR velocities
of Zeeman pairs by replacing the observed LSR velocities with the center LSR velocity of the pair.
Since spot position errors are not explicitly given in neither  \citet[][Tables~1~and~2]{Des98} nor  \citet[][Table~1]{Fis07a},
we employ the two following criteria to calculate them. First, using the formula of the Gaussian fit, position errors
are taken to be inversely proportional to the spot intensity.
Secondly, using the results of \citet[][Table~1]{Fis07a}, we request position errors
 to vary in the range 0.01~mas (for the strongest spots)
up to 1~mas (for the weakest spots).
Relative proper motions are determined by comparing the positions of the emission peaks of persistent spots between the
1994 May and 2006 June EVN epochs (the linear-fit method). Measuring proper motions over a time baseline of about 12~yr, 
the average uncertainty in OH maser positions of a few tenths of milliarcsec, at the \T\ distance of 1.95~kpc, 
translates into a velocity uncertainty of a few tenths of kilometer per second.

  \begin{figure*}
   \centering
   \includegraphics[width=\textwidth,angle=-90]{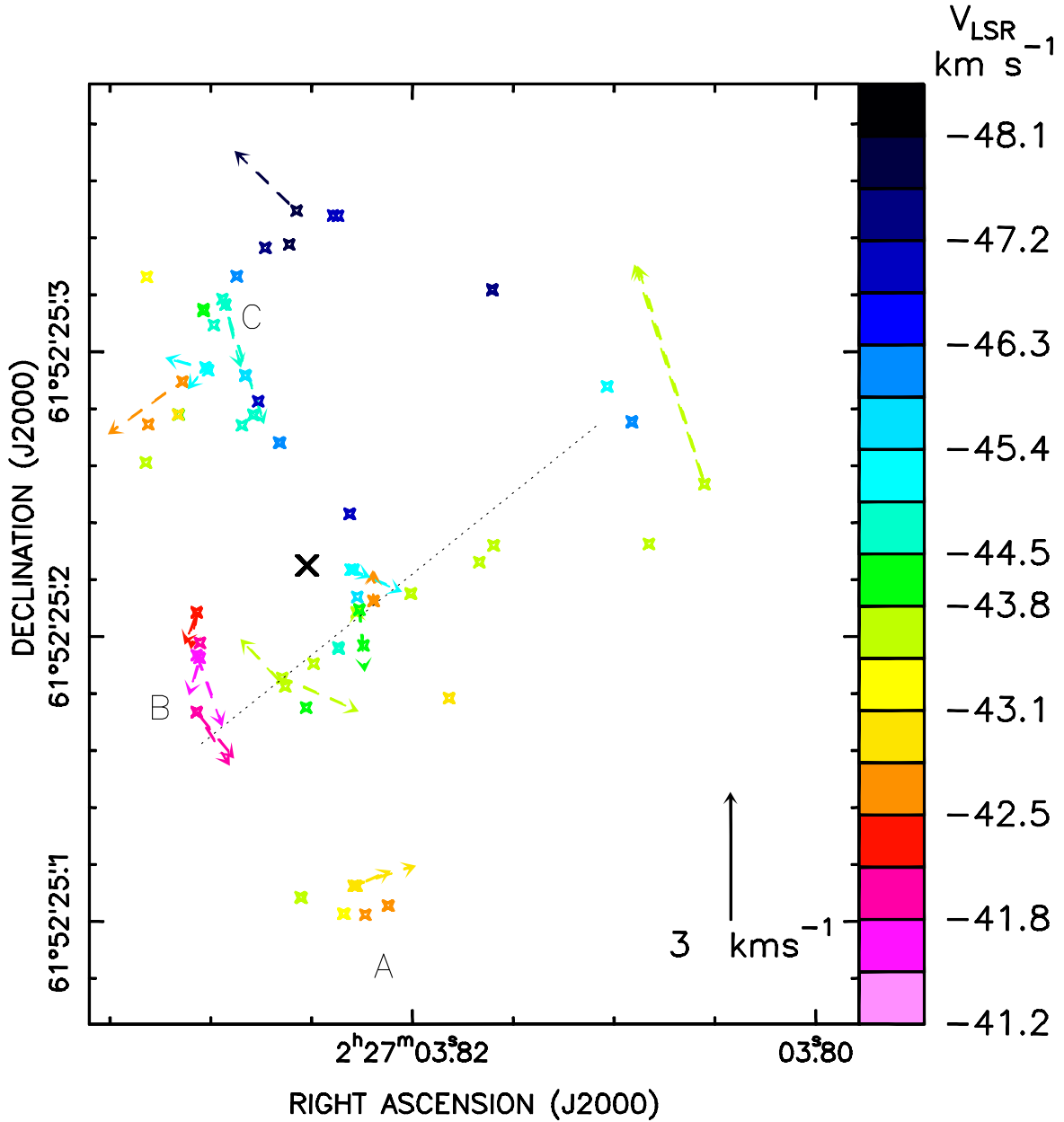}
      \caption{Absolute position and proper motions of OH 6035~MHz masers in the northern clump of \T.
      \emph{Stars} give absolute spot positions.
      \emph{Colors} are used to indicate the maser LSR velocities, according to the colored scale on the right-hand side
       of the plot, with \emph{green} denoting the  mean LSR velocity, at $-44.1$~\kms, of persistent \M\ 12~GHz masers.
        \emph{Dashed arrows} indicate proper motions relative to the ``center of motion'' of maser spots, whose position is marked with a \emph{black cross}.
       The proper motion amplitude scale is given at the bottom right of the plot. The \emph{dotted line} gives the approximate location and orientation of
        the ``linear structure at P.A. = 130\degr--140\degr'' drawn by both OH 6035~MHz and \M\ 12~GHz masers. 
        Capital letters ``A'', ``B'' and ``C'' are used to label three
        associations of OH 6035~MHz and \M\ 12~GHz masers.
              }
         \label{oh6035_abs}
   \end{figure*}

  \begin{figure*}
   \centering
   \includegraphics[width=\textwidth,angle=-90]{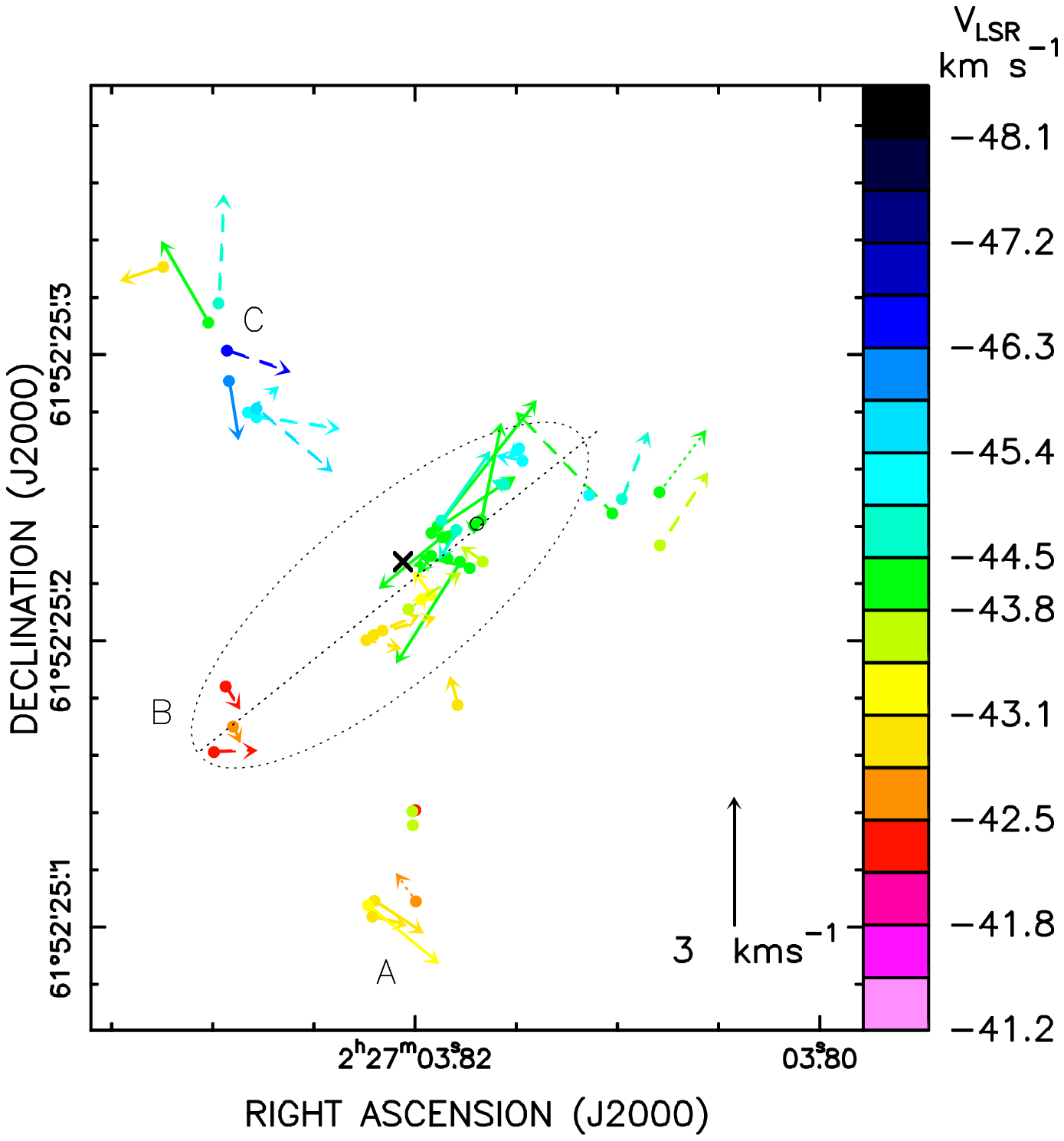}
      \caption{\small Absolute position and proper motions of \M\ 12~GHz masers in the northern clump of \T.
      \emph{Dots} give absolute spot positions.
      \emph{Colors} are used to indicate the maser LSR velocities, according to the colored scale on the right-hand side
       of the plot, with \emph{green} denoting the  mean LSR velocity, at $-44.1$~\kms, of persistent \M\ 12~GHz masers.
       \emph{Arrows} indicate proper motions relative to the ``center of motion'' of maser spots, whose position is marked with a \emph{black cross}.
      \emph{Dashed} and \emph{solid arrows} denote proper
      motions measured using two or more-than-two VLBA epochs, respectively.  \emph{Dotted arrows} are used for the most uncertain measurements.
      The proper motion amplitude scale is given at the bottom right of the plot.
       The \emph{dotted line} gives the approximate location and orientation of
        the ``linear structure at P.A. = 130\degr--140\degr''  drawn by both OH 6035~MHz and \M\ 12~GHz masers,
      and the \emph{dotted ellipse} embraces the 12~GHz maser spots belonging to the ``linear structure''.
      The \emph{small circle} denotes the position on the sky of the center of the rotating disk
      which we have fitted to the motion of the 12~GHz masers in the ``linear structure'' (see Sect.~\ref{mod_kin}).
       Capital letters ``A'', ``B'' and ``C'' are used to label three
        associations of OH 6035~MHz and \M\ 12~GHz masers.
              }
         \label{ch3oh_abs}
   \end{figure*}

In the northern clump, we find that 6031~MHz OH masers present
a close correspondence of positions and LSR velocities with a subset of 6035~MHz masers,
and the relative proper motions of nearby spots of the two OH emissions are also similar. Since it appears that
the two OH maser lines trace essentially the same motion, for the purpose of comparing OH and \M\ maser
kinematics, we employ the OH 6035~MHz masers, which
are more intense and numerous than the 6031~MHz masers.
Figures~\ref{oh6035_abs}~and~\ref{ch3oh_abs} show velocities and absolute positions of OH 6035 MHz and \M\ 12~GHz 
masers in the northern clump of \T, respectively.

Absolute positions of 12~GHz \M\ masers have been measured by XRZM with an accuracy of about 1~mas.
OH 6035~MHz maser absolute position has been established by \citet{Fis07a} with an error of about 10~mas.
In Fig.~\ref{oh6035_abs}, we have shifted the OH 6035~MHz masers with respect to their nominal position
by approximately 20~mas both to the west and to the south,
to overlap the ``linear structure at P.A. = 130\degr--140\degr'' drawn by both OH 6035~MHz and \M\ 12~GHz masers \citep[][MMWR1]{Des98}.
We are confident that
this shift improves the relative position between the two maser emissions, since it brings other groups of spots of the two
maser types, with similar line-of-sight and sky-projected velocities, to fall close each other in the sky.

Proper motions presented in Figs.~\ref{oh6035_abs}~and~\ref{ch3oh_abs}
are relative to the ``center of motion" of maser spots of the northern clump.
At each VLBI epoch, the position of the ``center of motion"  is defined by taking
the unweighted mean position of the spots persistent over all the observed epochs.  Calculating maser velocities relative to this point
is the same as first deriving proper motions relative to a given spot
and then subtracting the unweighted mean of all proper motions from each spot proper motion.
It is to be noted that the resulting proper motions do not depend on the choice
of the particular spot relative to which proper motions have been first derived.
As far as the masing gas distributes and moves symmetrically about the exciting star, and a sufficiently high number of maser spots are detected,
the ``center of motion" approaches more and more the star position and velocities relative to it resemble better and better the velocities
relative to the star.  Plotting velocities relative to the ``center of motion" of maser spots,
Figs.~\ref{oh6035_abs}~and~\ref{ch3oh_abs} permit a direct comparison of the kinematics of OH 6035~MHz
and \M\ 12~GHz masers in the northern clump of \T.
As the motion of the 12~GHz maser reference spot (\#~23) relative to the ``center of motion"  is small
(V$_{\mathrm{X}}$ = $0.61\pm0.06$~\kms, V$_{\mathrm{Y}}$ = $0.12\pm0.06$~\kms),
proper motions of the 12~GHz masers in the northern clump shown in Fig.~\ref{ch3oh_abs}
are similar to those of Fig.~\ref{prmot_rel_N}.

%

\subsection{Comparison of OH 6035~MHz and \M\ 12~GHz maser kinematics}

\label{comp_kin}

Looking at Figs.~\ref{oh6035_abs}~and~\ref{ch3oh_abs} one notes that the LSR velocities
of both OH 6035~MHz and \M\ 12~GHz masers are higher towards the southeast and
decrease moving towards the north-northwest. At the centre of the plotted field of view,
the ``linear structure at P.A. = 130\degr--140\degr'' is traced by both maser types over a total angular extent of about 150~mas.
It should be noted that two maser emissions appear to sample different segments of the line,
with OH 6035~MHz masers distributed towards the southeast and  \M\ 12~GHz masers mainly concentrated to the northwest.
Whereas 12~GHz masers belonging to the ``linear structure'' (inside the dotted ellipse in Fig.~\ref{ch3oh_abs}) present
a very regular variation of LSR velocities with position (monotonously decreasing towards the northwest), OH 6035~MHz masers show
a more scattered pattern of LSR velocities.

We identify at least three maser clusters
 (labeled in Figs.~\ref{oh6035_abs}~and~\ref{ch3oh_abs} with capital letters ``A", ``B" and ``C")
where spots of the two maser species are observed to be close in positions and with similar LSR velocities.
In clusters ``A" and ``B", proper motions of both OH 6035~MHz and \M\ 12~GHz masers present some degree of collimation
and the average direction of motion of spots of both maser types is also similar.
That, taken together with the good correspondence of the LSR velocities, suggests that the two maser emissions
in clusters ``A" and ``B" can partake in the same gas motion.
While most of the OH 6035~MHz masers of cluster ``C'' distribute along a southeast-northwest direction (about parallel to the
``linear structure at P.A. = 130\degr--140\degr''), with also a somewhat regular variation of LSR velocities (decreasing towards the northwest),
\M\ 12~GHz masers are observed only towards the center of the cluster ``C'' and emit across a narrower range of LSR velocities than the OH masers.
In this cluster, differently from what observed in clusters ``A" and ``B", the proper motions of both OH 6035~MHz
and \M\ 12~GHz masers are scattered over a large range of position angles.
The motion of the masing gas in cluster ``C'' appears more turbolent than in clusters ``A'' and ``B''.

Comparing the overall distribution of positions and velocities of the two maser emissions in the northern clump,
it seems that OH~6035~MHz and \M\ 12~GHz masers \emph{complement each other},
emerging from nearby (but likely distinct) cloudlets of masing gas with, in general, a rather smooth
variation of line-of-sight and sky-projected velocities.

\section{Revised model of \M\ 12~GHz maser kinematics in the northern clump}

\label{mod_kin}

\begin{figure}
 \centering
   \includegraphics[width=9cm,angle=0]{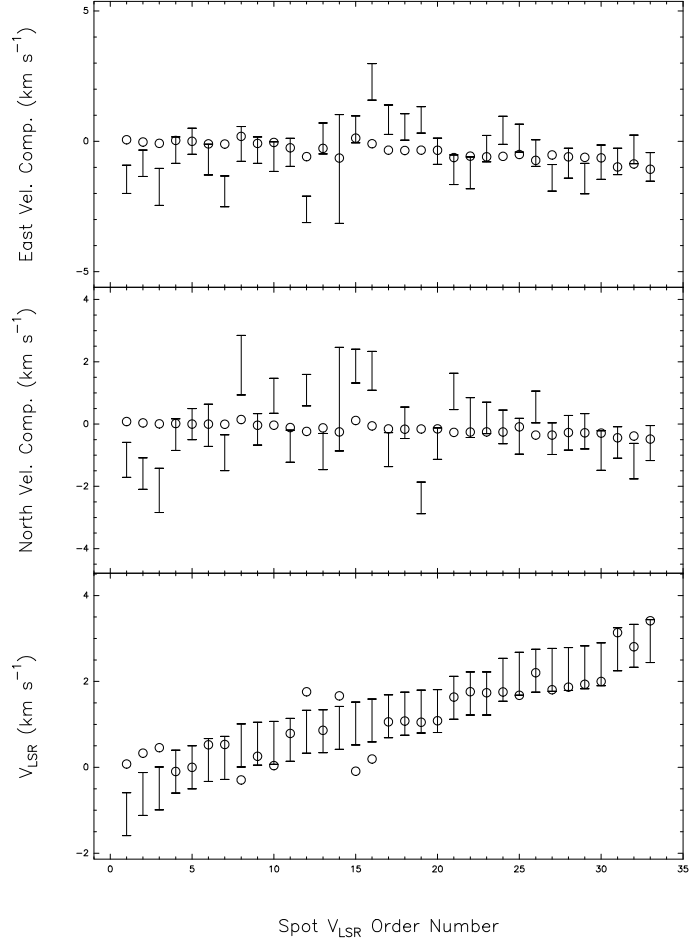}
      \caption{Upper, middle and lower panels compare the observed and model maser velocity components
       towards the east, the north and along the line of sight, respectively. In each panel, {\it errorbars} give
       values and estimated errors
       of the observed velocity components (relative to the 12~GHz maser reference spot \#~23),
       and {\it empty dots} indicate model values.  
    In the model fit calculation, to make the sky-projected and line-of-sight velocity       
       components weight the same, the velocity uncertainty $\Delta V$ used in the $\chi^2$ formula (see Equation~2 of MMWR2) is calculated
      as the quadratic sum of the measurement error and a constant term of 0.5~\kms. 
   Maser spots have been ordered by increasing LSR velocities, and the horizontal axis of each panel
       reports the spot order number.
              }        \label{mod_oldfit}
 \end{figure}

MMWR2 fit positions and velocities of all the detected \M\ 12~GHz masers in the northern clump using a model of conical flow,
where gas moves on the surface of a cone with a velocity resulting from the combination of an Hubble flow (directed along the cone generators)
and a rotation about the cone axis. Since the 12~GHz maser proper motions derived by MMWR2
(using the 1994 February and 1998 December VLBA epochs only) have large errors,
the conical flow model was constrained merely by 12~GHz maser LSR velocities (MMWR2, Fig.~6).
We have now tested the conical flow model with the new, more accurate measurements
of 12~GHz maser proper motions (see Table~\ref{prmot}). 
Using the same set of input maser spots (MMWR2, Table~2), we have looked for a
best-fit solution searching across a parameter range about the best-fit values determined in MMWR2.
With regard to the model geometry, the best-fit solution we have determined has 
both the sky-projected position of the cone vertex (offset less than 10~mas) 
 and the P.A. of the cone axis (differing less than 10$\degr$) close to the  MMWR2 best-fit values; the main difference
concerns the inclination angle of the cone axis and the cone opening angle, which in the newly fitted solution are required to be 
significantly larger (in the range \ 30$\degr$--35$\degr$) than the previous MMWR2 best-fit values ($<$10$\degr$).
With regard to the velocity field, the new solution comes with the coefficient of the Hubble flow about the double 
as the MMWR2 best-fit value, while the angular velocity of the rotation around the cone axis is obtained significantly smaller (by a factor $\approx$4)
than the previous best-fit value.  
However, our test indicates that the conical flow model is \emph{not able} to reproduce properly the improved values of sky-projected velocities.
Figure~\ref{mod_oldfit} compares the observed velocity components (towards the east, the north and along the line of sight) 
with the model values. 
While for the line-of-sight velocities the error-weighted rms residual is comparable with the meaurement errors (0.5~\kms),  
for both the velocity component towards the east and the north the error-weighted rms residual is of 1.0~\kms\  
and significantly exceeds the measurement errors (lower than 0.3~\kms) of most sky-projected velocities.
Thus, we are pushed to discard the conical flow model.

Fig.~\ref{ch3oh_abs} shows that
the velocity distribution of \M\ 12~GHz masers across the northern clump is quite complex.
Differently from what tempted in MMWR2, we restrain now from fitting a model to reproduce
positions and velocities of \emph{all} the 12~GHz maser spots of the northern clump.
We focus on the subset of 12~GHz maser spots in the northern clump  belonging to the ``linear structure at P.A. = 130\degr--140\degr'',
whose regular variation of LSR velocities with position presents evidence for some ordered motion.
In the following we discuss a model which reproduces positions and velocities of this group of masers only, identified
in Table~\ref{prmot} by the ``Component'' labels given in boldface characters.
The goodness of the fit of both the line-of-sight and sky-projected velocities, persuades us to present this alternative model.

We have fitted positions and 3-dimensional velocities of the linear distribution of 12~GHz masers with a rotating, flat disk model.
In the following, the model geometry is described using cilindrical coordinates: let us indicate with $r$ (the radius) and $z$ (the height)
the distances from the disk center, transverse and parallel to the model symmetry axis, respectively.
The radial profile of a flat disk can be well reproduced with a funnel profile, defined by the
equation \ $z = a \ln(r/r_{\mathrm{m}})$  with $r \ge r_{\mathrm{m}}$ ($r_{\mathrm{m}}$ being the minimum allowed distance from
the funnel axis), choosing a suitably small value for the funnel curvature coefficient $a$.
The velocity field consists of only the azimuthal component $v_{\phi} = v_m \, (r / r_{\mathrm{m}})^{\beta}$,
taken to be proportional to the power of the radial distance, with $v_m$, $r_{\mathrm{m}}$ and $\beta$ constants.
This expression
of the azimuthal velocity component is general enough to represent various, physicallly relevant motions.
Values of $\beta = -0.5$, 0 and 1 correspond to Keplerian, constant-velocity  and rigid
rotation, respectively.

Free parameters of the model are:  the position offset of the disk center (relative to the reference spot \#~23; see Table~\ref{prmot})
 towards the east, X$_{\mathrm{c}}$, and towards the north, Y$_{\mathrm{c}}$;
the position angle on the sky $p_{\mathrm{a}}$ (measured from north to east)  and
the inclination angle with the line of sight $i_{\mathrm{a}}$ of the disk axis; the funnel curvature coefficient $a$;
the velocity $v_m$ (at the minimum allowed distance $r_{\mathrm{m}}$)  and the exponent $\beta$
describing the power-law dependence of the velocity on the radial distance. $r_{\mathrm{m}}$, the minimum allowed distance from the disk axis,
is fixed to the value of 1~mas, negligible with respect to the size of the 12~GHz maser linear distribution (about 150~mas)
so to be irrelevant in the determination of the model geometry. Appendix~\ref{appe} describes the details of the model fit.

\clearpage
\setcounter{table}{2}
\begin{table*}
\caption{Best-fit parameters of the flat disk model for the 12~GHz maser ``linear structure''}             
\label{mod_par}      
\begin{tabular}{c c c c c c c}        
\hline\hline                 
\noalign{\smallskip}
\multicolumn{2}{c}{Disk Center Position} & \multicolumn{2}{c}{Disk Axis Orientation} & Curvature Coefficient & \multicolumn{2}{c}{Velocity Field} \\
X$_{\mathrm{c}}$ & Y$_{\mathrm{c}}$  & $p_{\mathrm{a}}$ & $i_{\mathrm{a}}$ &  $a$
& $\beta$ & $v_m$  \\    
(mas)     &  (mas)      &  (deg) & (deg) &   &  & (\kms)  \\
\noalign{\smallskip}
\hline                        
\noalign{\smallskip}
   16$\pm$5 & $-22\pm$5  & 41$\pm$6 & 103$\pm$6  & 0.003$\pm$0.001 & 0.5$\pm$0.5 &  $-0.19\pm$0.06  \\      
\noalign{\smallskip}
\hline   
\noalign{\smallskip}
\end{tabular}
\begin{flushleft}
{ \footnotesize  Note.-- Columns~1~and~2 give the position offsets
of the disk center from the reference spot \#~23 (see Table~\ref{prmot})
towards the east and north, respectively; Columns~3~and~4 report the P.A. (measured
from north to east) and the inclination angle (with respect to the line of sight) of the disk axis, respectively;
Column~5 gives the curvature coefficient of the funnel surface used to represent the flat disk;
Columns~6~and~7 give the parameters
of the power-law function expressing the dependence of the velocity on the radial distance, that is
the power-law exponent and the velocity at the minimum allowed distance, respectively.
The minimum allowed radius is fixed to the value of 1~mas. 
}
\end{flushleft}
\end{table*}

 \begin{figure}
  \centering
   \includegraphics[width=9cm,angle=0]{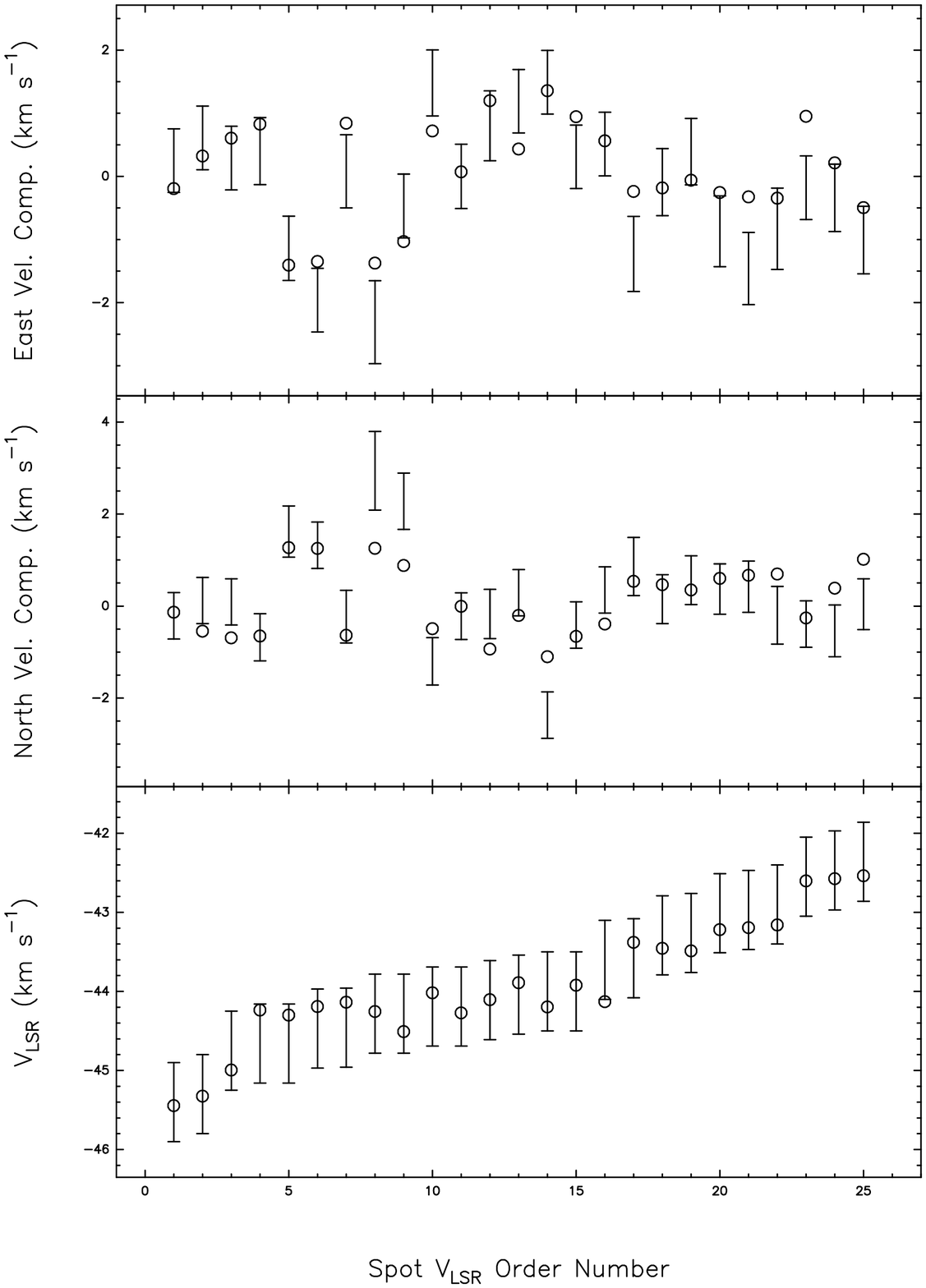}
      \caption{Upper, middle and lower panels compare the observed and model maser velocity components
       towards the east, the north and along the line of sight, respectively. In each panel, {\it errorbars} give
       values and estimated errors of the observed velocity components (relative to the ``center of motion''),
       and {\it empty dots} indicate model values.  Maser spots have been ordered by increasing LSR velocities, and the horizontal axis of each panel
       reports the spot order number.
              }        \label{mod_comp}
 \end{figure}

 \begin{figure*}
 \vspace{-2.5cm}
   \includegraphics[width=17.5cm,angle=-90]{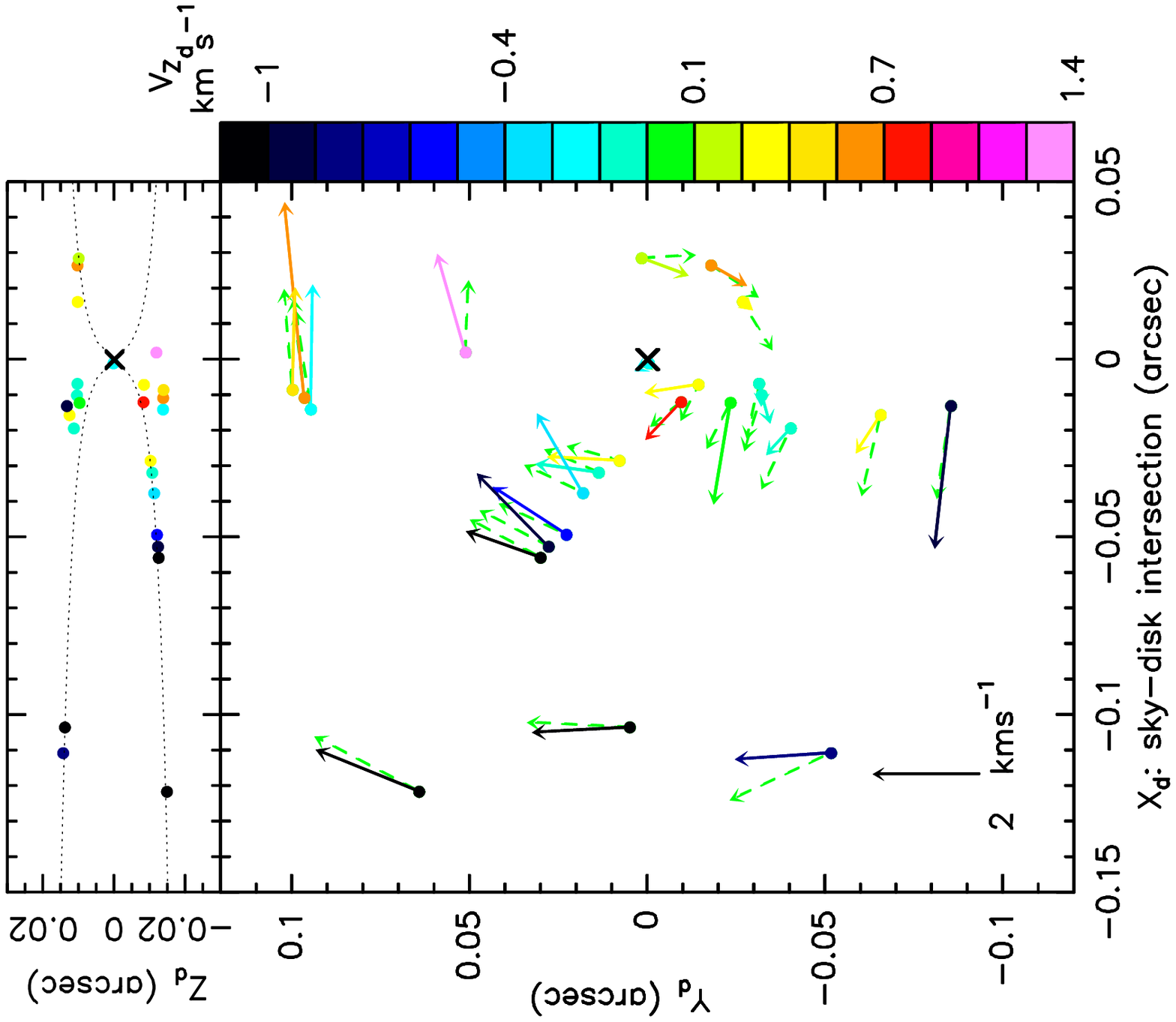}
      \caption{\small {\bf Lower panel:} Model maser positions projected onto the best-fit disk midplane and comparison of the model
    and the observed velocities.  The labels X$_\mathrm{d}$,  Y$_\mathrm{d}$  and Z$_\mathrm{d}$ indicate the axes
    (and the corresponding coordinates) of a Cartesian reference system,
   where  Z$_\mathrm{d}$ is the disk axis (P.A. = $p_{\mathrm{a}}$ = 41\degr, $i$ = $i_{\mathrm{a}}$ = 103\degr),
   X$_\mathrm{d}$ is the intersection of the disk midplane with the sky (oriented at P.A. = -49\degr and positive to NW), 
  and Y$_\mathrm{d}$ is the direction on the disk midplane perpendicular to X$_\mathrm{d}$ \ (with the Y$_\mathrm{d}$ positive semi-axis oriented so 
  to make \  (X$_\mathrm{d}$,Y$_\mathrm{d}$, Z$_\mathrm{d}$) \ a right-handed reference system).
      \emph{Dots} give model spot positions and
      \emph{colors} are used to indicate maser velocities projected along the disk axis, according to the colored scale on the right-hand side
       of the plot, with \emph{green} denoting the null velocity.
       \emph{Solid arrows} give observed velocities (relative to the ``center of motion'') projected onto the disk midplane,
      whereas \emph{dotted arrows} represent model velocities.
      The proper motion amplitude scale is given at the bottom of the plot.
      The disk center position is marked with a \emph{black cross}.
      \ {\bf Upper panel:} Model maser positions projected onto the  \ X$_\mathrm{d}$--Z$_\mathrm{d}$ plane. \emph{Dots},
      \emph{colors} and the \emph{black cross} have the same meaning as in the lower panel. The \emph{dotted lines}
      indicate the funnel profile used to reproduce the radial profile of the best-fit disk. Note that a spot  
       with \ Y$_\mathrm{d} \neq 0$ \  has a radial distance larger than its coordinate \ X$_\mathrm{d}$ \ 
      and consequently its Z$_\mathrm{d}$ coordinate  (in absolute value) is  always higher than that of the funnel profile
      evaluated at  the \ X$_\mathrm{d}$ coordinate of the spot.
              }        \label{mod_disk}
 \end{figure*}

Table~\ref{mod_par} lists the best-fit parameters of the disk model, corresponding to a minimum value of $\chi^2 = 1$ (see Appendix~\ref{appe}).
The error of each parameter has been evaluated with the offset from the best-fit value which causes a significative increase (by 20-40\%) of the $\chi^2$ value.
Figure~\ref{mod_comp} compares the observed velocity components (towards the east, the north and along the line of sight) with the fit values.
The error-weighted rms fit residual is 0.6, 0.7 and 0.2~\kms for the velocity component towards the east, the north and along the line of sight,
respectively. The ``center of motion'', relative to which maser velocities are derived, is found to be almost at rest in the model reference frame.
Following the procedure described in the Appendix~\ref{appe}, we have verified that the bias affecting the derived maser velocities 
 is small: $-0.2$, 0.0 and $-0.4$~\kms,  for the velocity component towards the east, the north and along the line of sight, respectively.

Figure~\ref{mod_disk} shows the maser positions on the best-fit disk and compares the model and the observed velocities.
Let us indicate with \ X$_\mathrm{d}$,  Y$_\mathrm{d}$  and Z$_\mathrm{d}$ the axes (and the corresponding coordinates)
 of a Cartesian reference system,
where  Z$_\mathrm{d}$ is the disk axis (P.A. = $p_{\mathrm{a}}$ = 41\degr, $i$ = $i_{\mathrm{a}}$ = 103\degr),
   X$_\mathrm{d}$ is the intersection of the disk midplane with the sky (oriented at P.A. = -49\degr and positive to NW), 
  and Y$_\mathrm{d}$ is the direction on the disk midplane perpendicular to  X$_\mathrm{d}$ 
\ (with the Y$_\mathrm{d}$ positive semi-axis oriented so 
  to make \ (X$_\mathrm{d}$,Y$_\mathrm{d}$, Z$_\mathrm{d}$) \ a right-handed reference system).
The rms difference of the observed and model velocity components is 0.7, 0.3 and 0.6~\kms
along the X$_\mathrm{d}$, Y$_\mathrm{d}$ and Z$_\mathrm{d}$ axis, respectively.
The upper panel of Figure~\ref{mod_disk} shows that the model maser distribution is significantly flat, the ratio
between the maximum radial distance ($\approx$140~mas) and the maximum height from the disk midplane ($\approx$15~mas)
being about 10. Except for one spot close to the disk center, \M\ 12~GHz masers are predicted to lay at distances of 10--15~mas from the
disk midplane. Note that, while the
distribution of model positions is about symmetrical along the Y$_\mathrm{d}$  and Z$_\mathrm{d}$ axes,
most spots are found at negative values of the coordinate X$_\mathrm{d}$.

The model predicts a square root dependence of the rotation velocity on the radius ($v_{\phi} \propto \sqrt{r}$),
which corresponds to a disk in centrifugal equilibrium with constant column density. In such a case one gets 
\ $v_{\phi} = \sqrt{\mathrm{G} \pi \Sigma_{0} \, r }$, where $\mathrm{G}$ is the Gravitational constant and
$\Sigma_{0}$ is the constant column density of the disk. Using the best-fit value of  \ $v_m = -0.19$~\kms\
(the velocity at the minimum allowed radial distance $r_{\mathrm{m}} = 1$~mas),
and the XRZM distance to \T\ of $1.95$~kpc (to convert angular to linear distances), one derives  \ $\Sigma_{0} = 60$~g~cm$^{-2}$. 
Our best-fit model suggests that \M\ 12~GHz maser emission
originates in a layer at an average height from the disk midplane of
$\approx$25~AU. In agreement with excitation models of \M\ 12~GHz masers \citep{Cra05} (requiring an H$_2$ numerical density of the order of
10$^{7}$--10$^{8}$~cm$^{-3}$), the mass density of the masing gas should be of the order of \ 10$^{-16}$~g~cm$^{-3}$.
To match the derived value of $\Sigma_{0}$, close to the disk midplane the gas
density has to increase by several order of magnitudes above the value estimated for the maser layer.
Let us indicate with $\rho$~[g~cm$^{-3}$] the disk mass density
and assume that it decreases exponentially with the height from the disk midplane:
$ \rho = \rho_0 \exp{(-|\mathrm{Z_d}| / \mathrm{H_d})}$, where $ \rho_0$ and H$_\mathrm{d}$ are the disk midplane mass density and
height scale, respectively. The knowledge of the approximate position and density of the maser layer, and of the
disk column density  \ $\Sigma_{0}$, allows us to determine the values of $ \rho_0$ and  H$_\mathrm{d}$.
We find H$_\mathrm{d}  \approx  3$~AU and $\rho_0 \sim 5 \times 10^{-13}$~g~cm$^{-3}$
(corresponding to \  n$_{\mathrm{H_2}} \sim 10^{11}$~cm$^{-3}$). Inside the volume sampled by the \M\ masers
(within a radius of 140~mas and an height from the disk midplane of 15~mas),
the disk mass is estimated  to be of  \ $\sim$1.5~M$_{\sun}$.

\subsection{Model Discussion}

\label{mod_discu}

Positions and velocities of \M\ 12~GHz masers
belonging to the ``linear structure'' in the \T\ northern clump
are well fitted with a flat disk geometry and a pure rotation field, which might indicate
that this subset of 12~GHz maser spots is tracing a self-gravitating structure.
The  ``linear structure'' drawn by the 12~GHz masers is located on the sky close
 to the peak of high frequency
  (15--23~GHz) continuum images of \T\ (see, for istance, MMWR2, Fig.5),
offset to the northwest by a few tenths of arcsec.
UV radiation excaping from  the \T\ UC H{\sc ii} region might photo-evaporate a nearby
circumstellar disk, and create a disk atmosphere sufficiently rich of gas-phase methanol,
and irradiated by a strong enough far-infrared field, to produce strong maser lines.
\M\ 12~GHz masers have never been observed in association with low-mass stars,
and if the ``linear structure'' of 12~GHz masers in the  \T\ northern clump
emerges from a low-mass disk, it is plausible that an \emph{external} radiation field is
responsible for the maser excitation.
Maser emission would be preferably excited on the disk side facing the UC H{\sc ii} region,
which would account for the asymmetrical distribution of 12~GHz maser spots about the
disk center predicted by the model.  Looking at Fig.~\ref{ch3oh_abs}, most of the 12~GHz masers
of the ``linear structure''
are distributed to the southeast of the disk center, along directions close to that pointing to the
peak of the radio continum images.

The comparison of Figs.~\ref{oh6035_abs}~and~\ref{ch3oh_abs} shows that the variation of LSR velocities
 of the OH 6035~MHz masers along the ``linear structure'' is less regular than for the \M\ 12~GHz masers.
The few proper motions of the 6035~MHz spots belonging to the ``linear structure'' are directed
approximately perpendicular to the elongation of the maser distribution  and cannot be reproduced with the simple rotation model
used for the 12~GHz masers. Our interpretation is that OH 6035~MHz masers can trace more external, less dense and warmer portions of the disk atmosphere,
where the H{\sc ii} region UV radiation can photo-dissociate heavier molecules and produce high abundances of gas-phase OH.
Note that, except for a few spots, all the OH 6035~MHz maser emission originates from the southeastern portion
of the ``linear structure'', that, in agreement with our model, would correspond to the (sky-projected) disk side facing the stronger source of external radiation.
More external portions of the disk atmosphere are likely more turbolent and gas could partly flow away from the disk.

In Sect.~\ref{comp_kin} we have discussed some properties
of  the OH 6035~MHz and \M\ 12~GHz maser association
labeled ``C'' in Figs.~\ref{oh6035_abs}~and~\ref{ch3oh_abs}.
The derived spot proper motions suggest that both maser emissions trace
outflowing/expanding motions from the central part of the maser cluster,
with speeds of a few \kms. We speculate that the maser cluster ``C'' might
originate from a low-mass circumstellar disk in a more evolved phase of photo-evaporation
than that harbouring the ``linear structure'' of 12~GHz masers discussed above.
Comparing with the ``linear structure'', the maser association ``C''
contains relatively few (and weak) \M\ masers with respect to OH masers.
That might indicate that the gas of maser cluster ``C'' has been more heavily processed by the H{\sc ii} region UV radiation.
The measured proper motions of the \M\ 12~GHz masers might
indicate that  the denser portions of the disk atmosphere are also expanding.

\section{Conclusions}
\label{conclu}

Employing seven VLBA epochs spanning a lapse of time of about 10~yr, 
this work determines accurate proper motions of the \M\ 12~GHz masers in the
northern clump of \T. The main features of the 12~GHz maser emission are persisting
and most of the strongest maser spots ($\ge$5~Jy) are detected at all the epochs.
The best measured sky-projected velocities have amplitudes in the
range \ 0.3--2~\kms\ and the corresponding uncertainties are 
as small as 0.1~\kms.

The conical flow model proposed by MMRW2 to interpret the kinematics of the
\M\ 12~GHz masers in the northern clump of \T, is tested with the new, more
accurate measurements of spot proper motions.  The result shows that the conical
flow model is not able to reproduce adequately the 12~GHz maser sky-projected
velocities and has to be discarded.
We consider a subset of 12~GHz masers in the northern clump belonging to the
``linear structure at P.A. = 130\degr--140\degr'',
whose regular variation of LSR velocities with position presents evidence for some ordered motion.
Positions and 3-dimensional velocities of this group of 12~GHz maser spots are well fitted
with a flat disk geometry and a pure rotation field.  This ``linear distribution''
of 12~GHz masers could trace a self-gravitating, low-mass circumstellar disk,
in the phase of being photo-evaporated by the strong UV-radiation field excaping from
 the \T\ UC H{\sc ii} region. 

Using literature data, we have derived proper motions of the OH 6.7~GHz masers
in the northern clump of \T. 
Comparing the overall distribution of positions and velocities of the OH and \M\ maser emissions in the northern clump,
it seems that OH~6035~MHz and \M\ 12~GHz masers complement each other,
emerging from nearby (but likely distinct) cloudlets of masing gas with, in general, a rather smooth
variation of line-of-sight and sky-projected velocities.

\begin{acknowledgements}
We would like to thank Riccardo Cesaroni for useful comments and careful reading of the article.
This work was supported by the Chinese NSF through grants NSF 10673024, NSF 10733030, NSF 10703010
and NSF 10621303, and NBRPC (973 Program) under grant 2007CB815403.
\end{acknowledgements}

\appendix

\section{Disk model fit}

\label{appe}

This section describes the fit of the flat disk model
to the ``linear distribution'' of 12~GHz masers in the northern clump.
Although the model has many (7) free parameters (see Sect.~\ref{mod_kin}), our observations constrain most of them
to vary within a narrow range of values.  The position of the disk center is searched inside an area of  \ $50 \times 50$~mas
to the northwest of the ``center of motion'' of the 12~GHz masers (see Fig.~\ref{ch3oh_abs}),
where most of the strongest 12~GHz masers of the ``linear structure'' concentrate.
As the maser distribution is significantly elongated (the ratio of the major onto
the minor axis of the ``linear structure'' is about 5), the disk should be seen close to edge-on.
Accordingly, the position angle of the disk axis, $p_{\mathrm{a}}$, is required to be close (within $\pm20\degr$) to the direction
(at P.A. = 45\degr) perpendicular to the maser line (at P.A. = 135\degr),
and the inclination angle of the disk axis (with respect to the line of sight), $i_{\mathrm{a}}$, is searched
over the range 70\degr--110\degr.
In order to obtain a flat and thin disk, the funnel curvature coefficient $a$ has to be $\ll1$ and we take it within the range \ $10^{-3}$--$10^{-2}$.
The power-law exponent $\beta$ is looked for in the range [$-1$,1], appropriate to reproduce typical velocity fields observed in molecular clumps and cores.
In correspondence with a given value of the exponent  \ $\beta$, $v_m$ (the velocity at the minimum allowed radial distance $r_{\mathrm{m}} =1$~mas) 
is varied over a range of values suitable
to reproduce the observed spread in line-of-sight and sky-projected velocities.

For a given set of input parameters, one can compute maser positions and velocity
vectors  and compare the model velocities with the observed
velocities.
The best-fit model parameters are determined by minimizing the $\chi^2$ expression:

\begin{equation}
\label{chi}
\chi^{2} = \sum_{n = 1}^{25} \sum_{i=1}^{3} \left[ \begin{array}{c} \frac{v_{i,n} - V_{i,n}}{\Delta V_{i,n}} \end{array} \right]^{2}
\end{equation}

where the index $i$ indicates the three (X, Y and Z) velocity
components (towards the east, towards the north and along the line of sight, respectively)
and the index $n$ refers to each of the 25 12~GHz maser spots belonging to the ``linear structure'',
for which the model is constructed. The lowercase \ $v$ \ denotes the velocity component
computed with the model, whereas the uppercase \ $V$ is the observed velocity
component (relative to the  ``center of motion'') with the corresponding uncertainty
given by $\Delta V$. Since the sky-projected velocities are determined
with respect to the ``center of motion'' (defined as described in Sect.~\ref{oh_pm}),
for consistency the systemic velocity of the disk is taken equal to the unweighted mean LSR velocity ($-44.1$~\kms)
of the persistent 12~GHz maser spots. If the ``center of motion'' has
a non negligible motion with respect to the star driving the maser kinematics,
the measured velocities differ from the ``true'' velocities by a constant vector.
To deal with that, 
after determining the best-fit parameters,  a least-square fit
of the model velocites versus the observed velocities is performed to estimate a possible offset for each of the three (observed) velocity components.
Then the $\chi^2$ value is recalculated after correcting the observed velocities for the derived offset.
This procedure allows one to estimate the velocity vector of the ``center of motion'' in the 
model reference frame without increasing the number of model free parameters.
The accuracy of most sky-projected velocities
(see Table~\ref{prmot}) is estimated to be significantly better than that
of the line-of-sight velocities, whose uncertainty is about 0.5~\kms, the 
velocity resolution of our VLBA data after hanning-smoothing. 
In the model fit calculation, to make the sky-projected and line-of-sight velocity
components weight the same,
the velocity uncertainty $\Delta V$ used in the $\chi^2$ formula is calculated
as the quadratic sum of the measurement error and a constant term of 0.5~\kms.

\bibliographystyle{aa}
\bibliography{biblio}

\clearpage

\setcounter{table}{1}

{\small
\begin{longtable}{cccrrrrcrrrr}
\caption{12~GHz Maser Proper Motions.} \\
\hline\hline
Component & V$_{\mathrm{LSR}}$ &  & \multicolumn{4}{c}{Relative Position} & & \multicolumn{4}{c}{Relative Velocity} \\
                    &                                       & & \multicolumn{1}{c}{X}   & \multicolumn{1}{c}{$\Delta$X} & \multicolumn{1}{c}{Y} & \multicolumn{1}{c}{$\Delta$Y} & & \multicolumn{1}{c}{V$_{\mathrm{X}}$} & \multicolumn{1}{c}{$\Delta$V$_{\mathrm{X}}$} & \multicolumn{1}{c}{V$_{\mathrm{Y}}$} & \multicolumn{1}{c}{$\Delta$V$_{\mathrm{Y}}$} \\
                    & (\kms)                             & & \multicolumn{2}{c}{(arcsec)} &  \multicolumn{2}{c}{(arcsec)} & & \multicolumn{2}{c}{(\kms)} &  \multicolumn{2}{c}{(\kms)} \\
\hline
\endfirsthead
\caption{continued.}\\
\hline\hline
Component & V$_{\mathrm{LSR}}$ &  & \multicolumn{4}{c}{Relative Position} & & \multicolumn{4}{c}{Relative Velocity} \\
                    &                                       & & \multicolumn{1}{c}{X}   & \multicolumn{1}{c}{$\Delta$X} & \multicolumn{1}{c}{Y} & \multicolumn{1}{c}{$\Delta$Y} & & \multicolumn{1}{c}{V$_{\mathrm{X}}$} & \multicolumn{1}{c}{$\Delta$V$_{\mathrm{X}}$} & \multicolumn{1}{c}{V$_{\mathrm{Y}}$} & \multicolumn{1}{c}{$\Delta$V$_{\mathrm{Y}}$} \\
                    & (\kms)                             & & \multicolumn{2}{c}{(arcsec)} &  \multicolumn{2}{c}{(arcsec)} & & \multicolumn{2}{c}{(\kms)} &  \multicolumn{2}{c}{(\kms)} \\
\hline
\endhead
\hline
\endfoot
 {\bf 1} & $ -42.5 $ & & $   0.10117 $ & $   0.00007 $ &  $  -0.09282 $ & $   0.00008 $ & & $  -0.77 $ & $  0.07 $ & $  -0.59 $ & $  0.07 $ \\
 \bf{  2 } & $ -42.5 $ & & $   0.10375 $ & $   0.00009 $ &  $  -0.07881 $ &  $ 0.00012 $  & &  $ \emph{ --0.31 } $ & $ \emph{  0.22 } $ & $  \emph{ --1.19  } $ & $ \emph{  0.28  } $  \\
   3 & $ -43.0 $ & & $   0.02270 $ & $   0.00010 $ &  $  -0.08531 $ & $   0.00011 $ & & $  -0.45 $ & $  0.09 $ & $   0.55 $ & $  0.10 $ \\
 \bf{  4 } & $ -43.0 $ & & $   0.04892 $ & $   0.00010 $ &  $  -0.05929 $ & $   0.00009 $ & & $ \emph{  --0.84 }  $ & $ \emph{  0.28 }  $ & $ \emph{  --0.28  } $ & $ \emph{  0.25 }  $ \\
   5 & $ -43.0 $ & & $   0.05246 $ & $   0.00009 $ &  $  -0.15921 $ & $   0.00009 $ & & $  -1.40 $ & $  0.08 $ & $  -0.47 $ & $  0.08 $ \\
 \bf{  6 } & $ -43.3 $ & & $   0.03216 $ & $   0.00008 $ &  $  -0.04681 $ & $   0.00008 $ & & $  \emph{  0.42  } $ & $ \emph{  0.20  } $ & $ \emph{  --0.09 }  $ & $ \emph{  0.20 }  $ \\
 \bf{  7 } & $ -43.3 $ & & $   0.03526 $ & $   0.00009 $ &  $  -0.04838 $ & $   0.00008 $ & & $ \emph{  --0.06  } $ & $ \emph{  0.22  } $ & $ \emph{  --0.50  } $ & $ \emph{  0.21  } $ \\
   8a & $ -43.1 $ & & $   0.05173 $ & $   0.00011 $ &  $  -0.15373 $ & $   0.00009 $ & & $  -1.77 $ & $  0.20 $ & $  -0.86 $ & $  0.24 $ \\
   8b & $ -43.1 $ & & $   0.05396 $ & $   0.00010 $ &  $  -0.15528 $ & $   0.00009 $ & & $  -2.44 $ & $  0.77 $ & $  -1.64 $ & $  0.44 $ \\
   9 & $ -43.7 $ & & $  -0.04806 $ & $   0.00010 $ &  $  -0.02947 $ & $   0.00011 $ & & $  \emph{ --1.09  } $ & $ \emph{  0.26  } $ & $ \emph{   1.05 }  $ & $ \emph{  0.30 }  $ \\
\bf{  10  } & $ -43.6 $ & & $   0.01391 $ & $   0.00009 $ &  $  -0.03518 $ & $   0.00008 $ & & $  -0.28 $ & $  0.08 $ & $   0.20 $ & $  0.07 $ \\
\bf{  11 }  & $ -44.0 $ & & $   0.01842 $ & $   0.00007 $ &  $  -0.03747 $ & $   0.00007 $ & & $   0.55 $ & $  0.07 $ & $   0.04 $ & $  0.07 $ \\
\bf{  12 }  & $ -44.1 $ & & $   0.02618 $ & $   0.00009 $ &  $  -0.03410 $ & $   0.00008 $ & & $   \emph{ 0.83  } $ & $ \emph{  0.26 }  $ & $ \emph{  --0.82 }  $ & $ \emph{  0.22  } $ \\
\bf{  13 }  & $ -44.0 $ & & $   0.03194 $ & $   0.00007 $ &  $  -0.03315 $ & $   0.00007 $ & & $  -0.38 $ & $  0.07 $ & $  -0.63 $ & $  0.07 $ \\
\bf{  14a  } & $ -44.3 $ & & $   0.01449 $ & $   0.00008 $ &  $  -0.02068 $ & $   0.00008 $ & & $  -1.01 $ & $  0.14 $ & $   2.05 $ & $  0.39 $ \\
\bf{  14b }  & $ -44.2 $ & & $   0.01693 $ & $   0.00007 $ &  $  -0.02254 $ & $   0.00007 $ & & $  -0.61 $ & $  0.10 $ & $  -0.51 $ & $  0.20 $ \\
 \bf{ 15 }  & $ -44.0 $ & & $   0.02181 $ & $   0.00009 $ &  $  -0.03522 $ & $   0.00008 $ & & $   0.82 $ & $  0.08 $ & $  -2.37 $ & $  0.08 $ \\
\bf{  16  } & $ -44.5 $ & & $   0.02558 $ & $   0.00014 $ &  $  -0.02628 $ & $   0.00012 $ & & $  \emph{  0.11 }  $ & $  \emph{ 0.32 }  $ & $ \emph{  --0.88  } $ & $ \emph{  0.29  } $ \\
\bf{  17  } & $ -44.5 $ & & $   0.03182 $ & $   0.00008 $ &  $  -0.02519 $ & $   0.00008 $ & & $  -2.61 $ & $  0.07 $ & $   1.09 $ & $  0.07 $ \\
  18 & $ -44.3 $ & & $   0.10981 $ & $   0.00016 $ &  $   0.04831 $ & $   0.00028 $ & & $   0.46 $ & $  0.13 $ & $   1.86 $ & $  0.21 $ \\
\bf{  19  } & $ -44.8 $ & & $   0.00639 $ & $   0.00007 $ &  $  -0.00829 $ & $   0.00007 $ & & $  -0.34 $ & $  0.07 $ & $  -0.17 $ & $  0.07 $ \\
 \bf{ 20  } & $ -44.7 $ & & $   0.02315 $ & $   0.00037 $ &  $  -0.02412 $ & $   0.00018 $ & & $  -0.42 $ & $  0.20 $ & $  -0.71 $ & $  0.13 $ \\
\bf{  21b } & $ -44.3 $ & & $   0.02961 $ & $   0.00008 $ &  $  -0.02279 $ & $   0.00008 $ & & $  -2.77 $ & $  0.50 $ & $   2.57 $ & $  0.73 $ \\
\bf{  21a  } & $ -44.7 $ & & $   0.02834 $ & $   0.00007 $ &  $  -0.02081 $ & $   0.00007 $ & & $  -1.77 $ & $  0.14 $ & $   1.28 $ & $  0.28 $ \\
  22 & $ -44.8 $ & & $   0.10625 $ & $   0.00018 $ &  $   0.05504 $ & $   0.00035 $ & & $ \emph{  --0.10 }  $ & $ \emph{  0.44  } $ & $  \emph{  1.89  } $ & $ \emph{  0.81 }  $ \\
\bf{  23  } & $ -45.3 $ & & $   0.0 $ & $   0.0 $ &  $   0.0 $ & $   0.0 $ & & $   0.0 $ & $  0.0 $ & $   0.0 $ & $  0.0 $ \\
 \bf{ 24  } & $ -45.4 $ & & $   0.00116 $ & $   0.00008 $ &  $   0.00438 $ & $   0.00008 $ & & $  -0.34 $ & $  0.07 $ & $  -0.34 $ & $  0.07 $ \\
  25 & $ -45.1 $ & & $   0.09295 $ & $   0.00012 $ &  $   0.01525 $ & $   0.00010 $ & & $ \emph{  --1.92 }  $ & $ \emph{  0.31 }  $ & $ \emph{  --0.92 }  $ & $ \emph{  0.28 }  $ \\
  26 & $ -45.1 $ & & $   0.09595 $ & $   0.00012 $ &  $   0.01695 $ & $   0.00017 $ & & $ \emph{  --0.70 }  $ & $ \emph{  0.31  } $ & $ \emph{  --0.04  } $ & $ \emph{  0.46 }  $ \\
  27 & $ -45.9 $ & & $   0.10259 $ & $   0.00007 $ &  $   0.02792 $ & $   0.00008 $ & & $  -0.84 $ & $  0.07 $ & $  -1.59 $ & $  0.07 $ \\
  28 & $ -46.4 $ & & $   0.10328 $ & $   0.00008 $ &  $   0.03852 $ & $   0.00011 $ & & $  \emph{ --1.46  } $ & $ \emph{  0.21  } $ & $ \emph{  --1.15 }  $ & $  \emph{ 0.26  } $ \\
\bf{  29  } & $ -42.4 $ & & $   0.10782 $ & $   0.00009 $ &  $  -0.10174 $ & $   0.00011 $ & & $ \emph{  --0.98 }  $ & $ \emph{   0.23 }  $ & $ \emph{  --0.61  } $ & $  \emph{ 0.26  } $ \\
  30 & $ -42.6 $ & & $   0.03724 $ & $   0.00104 $ &  $  -0.15389 $ & $   0.00121 $ & & $  \emph{  0.49  } $ & $ \emph{  2.02 }  $ & $ \emph{   0.01  } $ & $ \emph{  2.45 }  $ \\
 \bf{ 31  } & $ -43.0 $ & & $   0.05202 $ & $   0.00012 $ &  $  -0.06085 $ & $   0.00011 $ & & $ \emph{  --1.43  } $ & $ \emph{  0.30 }  $ & $ \emph{  --0.23  } $ & $ \emph{  0.27 }  $ \\
 \bf{ 32  } & $ -42.9 $ & & $   0.05458 $ & $   0.00020 $ &  $  -0.06263 $ & $   0.00019 $ & & $ \emph{  --0.80 }  $ & $ \emph{  0.42  } $ & $ \emph{  --0.85  } $ & $ \emph{  0.39  } $ \\
 \bf{ 33  } & $ -43.6 $ & & $   0.03990 $ & $   0.00012 $ &  $  -0.05182 $ & $   0.00015 $ & & $ \emph{  --1.21  } $ & $ \emph{  0.35  } $ & $ \emph{   0.21 }  $ & $ \emph{  0.40 }  $ \\
 \bf{ 34  } & $ -44.2 $ & & $   0.02801 $ & $   0.00024 $ &  $  -0.02673 $ & $   0.00016 $ & & $   0.78 $ & $  0.16 $ & $  -1.48 $ & $  0.12 $ \\
  35 & $ -44.4 $ & & $  -0.04787 $ & $   0.00062 $ &  $  -0.01094 $ & $   0.00043 $ & & $ \emph{  --1.06 }  $ & $ \emph{  2.02 }  $ & $ \emph{   0.80  } $ & $ \emph{  1.59 }  $ \\
  36 & $ -44.7 $ & & $  -0.03473 $ & $   0.00010 $ &  $  -0.01332 $ & $   0.00010 $ & & $ \emph{  --0.59  } $ & $ \emph{  0.27 }  $ & $  \emph{  0.91 }  $ & $ \emph{  0.26  } $ \\
  37 & $ -44.2 $ & & $  -0.03143 $ & $   0.00010 $ &  $  -0.01836 $ & $   0.00009 $ & & $ \emph{   2.28  } $ & $ \emph{  0.49 }  $ & $  \emph{  1.71  } $ & $ \emph{  0.37 }  $ \\
  39 & $ -45.8 $ & & $   0.09296 $ & $   0.00023 $ &  $   0.01832 $ & $   0.00022 $ & & $ \emph{  --1.75  } $ & $ \emph{  0.50  } $ & $ \emph{  --2.13  } $ & $ \emph{  0.50 }  $ \\
  40 & $ -43.0 $ & & $  -0.04246 $ & $   0.00008 $ &  $  -1.16120 $ & $   0.00008 $ & & $  -3.58 $ & $  0.08 $ & $   0.28 $ & $  0.08 $ \\
  41 & $ -43.4 $ & & $  -0.03647 $ & $   0.00008 $ &  $  -1.15088 $ & $   0.00008 $ & & $  -3.76 $ & $  0.08 $ & $  -0.18 $ & $  0.08 $ \\
  42 & $ -43.3 $ & & $  -0.03485 $ & $   0.00008 $ &  $  -1.16300 $ & $   0.00013 $ & & $ \emph{  --3.67 }  $ & $ \emph{  0.21  } $ & $  \emph{  0.32  } $ & $ \emph{  0.30  } $ \\
  43 & $ -43.6 $ & & $  -0.05358 $ & $   0.00012 $ &  $  -1.17243 $ & $   0.00014 $ & & $  -3.08 $ & $  0.14 $ & $  -0.10 $ & $  0.16 $ \\
  44 & $ -43.4 $ & & $  -0.03464 $ & $   0.00007 $ &  $  -1.15906 $ & $   0.00011 $ & & $  -3.88 $ & $  0.07 $ & $   1.92 $ & $  0.09 $ \\
  45 & $ -43.0 $ & & $  -0.04271 $ & $   0.00012 $ &  $  -1.15498 $ & $   0.00016 $ & & $ \emph{  --3.57 }  $ & $ \emph{  0.27  } $ & $  \emph{  0.65  } $ & $ \emph{  0.34 }  $ \\
  46 & $ -43.1 $ & & $   0.12556 $ & $   0.00044 $ &  $   0.06779 $ & $   0.00075 $ & & $   0.12 $ & $  0.19 $ & $  -0.39 $ & $  0.28 $ \\
  48 & $ -44.2 $ & & $  -0.15772 $ & $   0.00043 $ &  $  -0.68709 $ & $   0.00027 $ & & $ \emph{  --2.02  } $ & $  \emph{ 1.22  } $ & $  \emph{  0.01  } $ & $  \emph{ 0.95  } $ \\
\label{prmot}
\end{longtable}
}

{\footnotesize For each maser spot, Col.~1 lists the ``Component''
label; Col.~2 gives the maser V$_{\mathrm{LSR}}$;
Cols.~3--6 report the spot position relative to the reference spot \#~23: Cols.3--4 and Cols.5--6
give the position offsets toward the east and north, respectively, with the associated errors;
Cols.~7--10 report the spot proper motion relative to the reference spot \#~23: Cols.7--8 and Cols.9--10
give the velocity components toward the east and north, respectively, with the associated errors.
Maser spots observed during the first two VLBA epochs only, have more uncertain velocities and their proper motion
components are given in italic characters. Spots with ``Component'' label given in boldface characters belong to the
``linear structure at P.A. = 130\degr--140\degr'' observed in the northern clump of \T, and are those whose positions and velocities
are reproduced with the kinematic model discussed in Sect.~\ref{mod_kin}.}

\end{document}